\documentclass[aps,prl,showpacs,notitlepage,twocolumn,superscriptaddress,nofootinbib,preprintnumbers]{revtex4-2}
\usepackage{bbm}
\usepackage{mathrsfs}
\usepackage{epsfig}
\usepackage{soul,xcolor}
\usepackage{graphicx}
\usepackage{amsfonts}
\usepackage{amsthm}
\usepackage[figuresright]{rotating}
\usepackage{amssymb}
\usepackage{amsmath}
\usepackage{dcolumn}
\usepackage{physics}
\usepackage{float}
\usepackage{bm}
\usepackage{physics}
\usepackage{verbatim}
\usepackage{braket}
\usepackage[normalem]{ulem}
\usepackage[ruled,vlined,linesnumbered]{algorithm2e}
\usepackage{algorithmic}
\usepackage{setspace}
\usepackage{lipsum}
\usepackage[colorlinks,linkcolor=blue,anchorcolor=blue,citecolor=blue,urlcolor=blue]{hyperref}
\usepackage{mathtools}

\newtheorem{theorem}{Theorem}

\newtheorem{lemma}{Lemma}
\newtheorem{definition}{Definition}
\newtheorem{proposition}{Proposition}
\usepackage{caption}
\begin{document}

\title{Provably Efficient Adiabatic Learning for Quantum-Classical Dynamics}

\author{Changnan Peng}
\affiliation{Department of Physics, Massachusetts Institute of Technology, Cambridge, MA 02139, USA}
\author{Jin-Peng Liu}
\affiliation{Center for Theoretical Physics, Massachusetts Institute of Technology, Cambridge, MA 02139, USA}
\affiliation{Simons Institute and Department of Mathematics, University of California, Berkeley, CA 94704, USA}
\author{Gia-Wei Chern}
\affiliation{Department of Physics, University of Virginia, Charlottesville, VA 22904, USA}
\author{Di Luo}
\thanks{correspondence: diluo@mit.edu}
\affiliation{Center for Theoretical Physics, Massachusetts Institute of Technology, Cambridge, MA 02139, USA}
\affiliation{The NSF AI Institute for Artificial Intelligence and Fundamental Interactions}
\affiliation{Department of Physics, Harvard University, Cambridge, MA 02138, USA}

\date{\today}

\begin{abstract}
Quantum-classical hybrid dynamics is crucial for accurately simulating complex systems where both quantum and classical behaviors need to be considered. However, coupling between classical and quantum degrees of freedom and the exponential growth of the Hilbert space present significant challenges. Current machine learning approaches for predicting such dynamics, while promising, remain unknown in their error bounds, sample complexity, and generalizability. In this work, we establish a generic theoretical framework for analyzing quantum-classical adiabatic dynamics with learning algorithms. Based on quantum information theory, we develop a provably efficient adiabatic learning  (PEAL) algorithm with logarithmic system size sampling complexity and favorable time scaling properties. We benchmark PEAL on the Holstein model, and demonstrate its accuracy in predicting single-path dynamics and ensemble dynamics observables as well as transfer learning over a family of Hamiltonians. Our framework and algorithm open up new avenues for reliable and efficient learning of quantum-classical dynamics.
\end{abstract}

\maketitle

\emph{Introduction.}--- Efficient simulation of quantum-classical hybrid dynamics is crucial to multi-scale modelings of a wide range of physical systems, opening new avenues for advancements in material science, chemistry, and drug discovery by providing a more comprehensive understanding of molecular interactions~\cite{Kapral:1999quantumclassicalhybrid,Kapral:2006quantumclassicalhybrid,Lin:2007quantumclassicalhybrid,vanderKamp:2013quantumclassicalhybrid,Bauer:2016quantumclassicalhybrid,Melo:2018quantumclassicalhybrid,Unke:2024quantumclassicalhybrid}. A common hybrid dynamics approach relies on the adiabatic approximation, where two well-separated timescales of a system allow one to treat the slow dynamics classically while quantum calculations are used to solve the fast, often electronic, degrees of freedom that adiabatically follow the classical dynamics. A well-known example is the Born-Oppenheimer approximation which is widely used in {\em ab initio} molecular dynamics~\cite{Marx09}.
The significance of quantum-classical hybrid dynamics lies in its potential to revolutionize how we model and predict the behavior of complex systems, ultimately pushing the frontiers of both fundamental research and practical applications.

The simulations of quantum-classical dynamics, however, is computationally challenging due to not only an exponentially large Hilbert space of quantum sub-systems and repeated time-consuming quantum calculations at every time step, but also the nonlinear differential equation coupled both the quantum and the classical variables. In the past two decades, machine learning (ML) has emerged as a powerful tool in developing force fields and inter-atomic potentials for {\em ab initio} molecular dynamics~\cite{behler07,bartok10,li15,shapeev16,behler16,botu17,zhang18,mcgibbon17,suwa19,chmiela17,chmiela18,sauceda20}. ML force-field approaches have recently been generalized to enable large-scale dynamical simulations of condensed-matter lattice systems~\cite{zhang21,zhang22,zhang23,cheng23,cheng23b}. 
This approach leverages massive datasets of quantum mechanical results to train models that can predict the potential energy surfaces with high precision.  ML-enhanced force fields facilitate the simulation of large intricate systems by accurately capturing the essential quantum mechanical effects while maintaining computational efficiency. Despite intensive studies and wide applications of ML force field models over past decades, critical questions on the error bound, sample complexity and generalizability of the ML methods have remained unresolved. 

In this work, we establish a generic theoretical framework for analyzing quantum-classical adiabatic dynamics with learning algorithms. We start with the analysis of the approximately constant linear model, derive the error bounded condition for the non-linear model, and introduce the relaxation method to check the error bounded property for a generic model, which lays down a solid foundation for the reliability of learning algorithms in quantum-classical adiabatic dynamics. Inspired by the recent development of quantum information theoretic learning theory for quantum many-body systems~\cite{Preskill:2022science,Preskill:2024nature,huang2020predicting,onorati2023efficient,haah2024learning,huang2023learning,caro2022generalization,gibbs2024dynamical,fanizza2024learning,levy2024classical,caro2023out,bakshi2024structure,onorati2023provably,huang2023learning2}, we develop a provably efficient adiabatic learning (PEAL) algorithm for quantum-classical dynamics, which offers a sample complexity scaling logarithmically with the system size and favorable scaling of evolution time. We benchmark our algorithm on the Holstein model and demonstrate accurate prediction of the single path dynamics and ensemble dynamics observables, as well as transfer learning across different couplings between quantum and classical degrees of freedom.

\begin{figure}[t!]
\vspace{10pt}
\centering
\includegraphics[width=8.5cm]{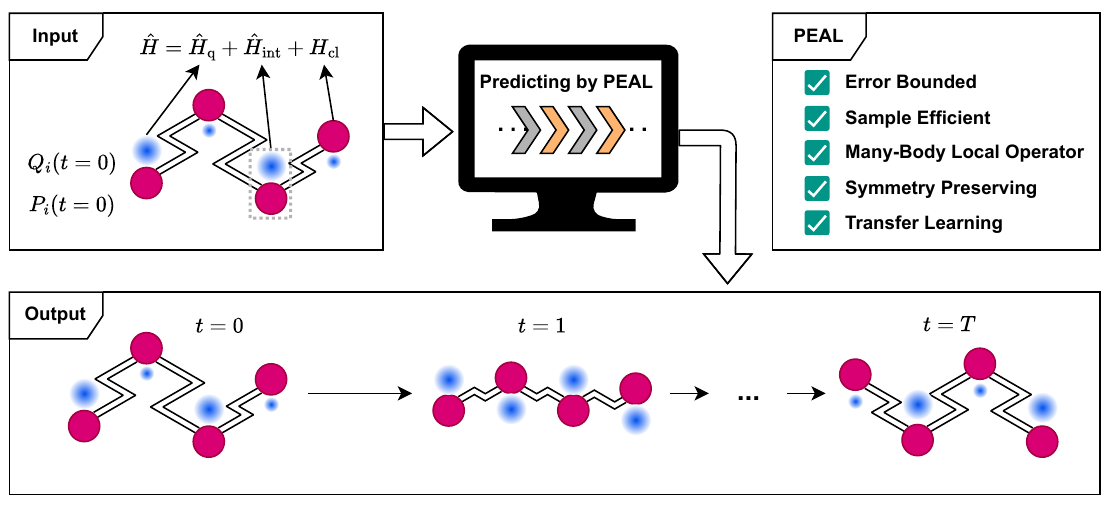}
\caption{Schematic diagram for PEAL.}
\vspace{-20pt}
\label{fig:cartoon}
\end{figure}

\emph{Adiabatic Quantum-Classical Dynamics Learning.}--- We consider a general quantum-classical Hamiltonian:

\begin{align}
    \hat{H} = \hat{H}_{q} + \sum_{\alpha,i} g_\alpha \hat{O}_{\alpha,i} G_{\alpha,i}(\Vec{P},\Vec{Q}) + H_{cl}(\Vec{P},\Vec{Q}),
    \label{eq:ham}
\end{align}
where $\hat{H}_{q}$ is the Hamiltonian operator for the quantum degrees of freedom, $g_\alpha$ is the quantum-classical coupling coefficient, $\alpha$ is the index of different types of couplings, $i\in\{1,\dots,L\}$ is the label of local regions (e.g. lattice sites), $\hat{O}_{\alpha,i}$ is the quantum operator that enters the coupling, $G_{\alpha,i}(\Vec{P},\Vec{Q})$ is a general function of the classical degree of freedom $\Vec{Q}=(Q_1,\dots,Q_L)$ and its canonical momentum $\Vec{P}=(P_1,\dots,P_L)$, and $H_{cl}(\Vec{P},\Vec{Q})$ is the classical Hamiltonian.

We study the adiabatic evolution dynamics driven by the quantum-classical Hamiltonian in Eq.~(\ref{eq:ham}). In this adiabatic limit, similar to the Born-Oppenheimer approximation in quantum molecular dynamics, the quantum subsystem is assumed to quickly relax to the ground state of the total Hamiltonian $\hat{H}$, while the classical degrees of freedom follow the Hamilton's equations of motion (EOM):
\begin{align}
    \frac{d}{dt} Q_j &= \sum_{\alpha,i} g_\alpha \langle\hat{O}_{\alpha,i}\rangle \frac{\partial}{\partial P_j} G_{\alpha,i} + \frac{\partial}{\partial P_j} H_{cl}, 
    \label{eq:eom_Q}
    \\
    \frac{d}{dt} P_j &= -\sum_{\alpha,i} g_\alpha \langle\hat{O}_{\alpha,i}\rangle \frac{\partial}{\partial Q_j} G_{\alpha,i} - \frac{\partial}{\partial Q_j} H_{cl} - \gamma P_j,
    \label{eq:eom_P}
\end{align}
where $j\in\{1,\dots,L\}$, $\langle\cdot\rangle$ is the ground state expectation, and $\gamma > 0$ is the damping coefficient due to dissipation.

The conventional way to solve the above equations is through an iterative scheme, where one first updates $\{Q_j(t), P_j(t)\}$ based on the classical EOM, and then solve the ground state from $\hat{H}$. The updated ground state is used to compute expectation values $\langle\hat{O}_{\alpha,i}\rangle$ which determine the driving terms of the EOM for the next step. However, repeated ground-state calculations of $\hat{H}$ at every time step could be computationally costly using quantum state solvers (QSS), such as exact diagonalization (ED)~\cite{ED:textbook}, density functional theory~\cite{ Hohenberg:1964DFT}, DMRG~\cite{white1993density}, neural network~\cite{luo2019backflow,carleo2017solving,luo2021gauge,luo2023gauge,luo2022autoregressive}, and quantum computers~\cite{wecker2015progress,mcclean2016theory,peruzzo2014variational,dorner2009optimal}  .

We consider a general quantum-classical ML model for learning such dynamics, which we call \textit{Adiabatic Dynamics Model Learning} (ADML). ADML consists of two components, which uses machine learning to predict the quantum observables and evolves the classical observables using classical numerical schemes. The ML force-field models widely used in quantum molecular dynamics can be viewed as special classes of ADML~\cite{behler07,bartok10,li15,shapeev16,behler16,botu17,zhang18,mcgibbon17,suwa19,chmiela17,chmiela18,sauceda20}. Our goal here is to predict the adiabatic dynamics of Eq.~(\ref{eq:ham}) with a learning-based approach. Given access to a dataset $\bigcup_{s=1}^{N_s}\{(g_\alpha, \Vec{P}(t), \Vec{Q}(t), \langle\hat{O}\rangle(t))_s : t\in\mathcal{T}_s\}$, where $\mathcal{T}_s$ is the set of sampled time steps, $N_s$ is the number of initial conditions $\{(g_\alpha, \Vec{P}(t_{\text{init}}), \Vec{Q}(t_{\text{init}}))_s\}_{s=1}^{N_s}$ sampled from a distribution $\mathcal{D}_{\text{init}}$, and $\hat{O}$ stands for $\hat{O}_{\alpha,i}$ in Eq.~(\ref{eq:ham}) or other operators of insterest but not in the Hamiltonian, the task is to design ADML for predicting the dynamics starting with other $\{ (g_\alpha, \Vec{P}(t_{\text{init}}), \Vec{Q}(t_{\text{init}}))\}$ from $\mathcal{D}_{\text{init}}$. 
In the following, we analyze and derive the error bounded conditions for ADML.

\textit{(i) Approximately constant linear model.} To serve as a starting point, we consider a simple example of Eq.~(\ref{eq:ham}),
$
    \hat{H} = \hat{H}_{q} + \sum_{i} g \hat{O}_{i} Q_{i} + \sum_{i} \left( \frac{1}{2M} P_{i}^2 + \frac{1}{2} k Q_{i}^2 \right),
$
where we only consider one type of quantum-classical coupling with $G_{\alpha,i}(\Vec{P},\Vec{Q})=Q_{i}$, and the classical Hamiltonian $H_{cl}(\Vec{P},\Vec{Q})$ is for simple harmonic oscillators with mass $M$ and spring constant $k$. Further assuming that during the dynamical process we are interested in, the response $\frac{\partial\langle\hat{O}_{i}\rangle}{\partial Q_i}$ is approximately a constant, and the off-diagonal response $\frac{\partial\langle\hat{O}_{i}\rangle}{\partial Q_j}$ ($i\neq j$) is approximately zero, the EOM of the system is then reduced to that of independent simple harmonic oscillators.

Focusing on the classical degree of freedom $Q_i$, we can view the quantum-classical coupling $g \hat{O}_{i} Q_{i}$ as a driving force on the oscillator. Since the EOM in Eq.~(\ref{eq:eom_Q}) and~(\ref{eq:eom_P}) are approximately linear in this case, the accumulated momentum and position errors between ML and the exact simulation using QSS, $p(t) = P_{i,\text{ML}}(t)-P_{i,\text{Exact}}(t)$, $q(t) = Q_{i,\text{ML}}(t)-Q_{i,\text{Exact}}(t)$ ($i$ index suppressed), also satisfy a similar EOM:
\begin{align}
    \frac{d}{dt} q(t) &= \frac{1}{M} p(t),
    \label{eq:eom_q}
    \\
    \frac{d}{dt} p(t) &= F(t) - K q(t) - \gamma p(t) + o(q(t))
    \label{eq:eom_p},
\end{align}
where we define the error force $F(t) = - g \delta\langle\hat{O}_i\rangle(t)$ and the error stiffness $K = g \frac{\partial\langle\hat{O}_i\rangle}{\partial Q_i} + k$. $\delta\langle\hat{O}_i\rangle(t) = \langle\hat{O}_{i}\rangle_{\text{ML}}(\Vec{Q}_{\text{ML}}(t)) - \langle\hat{O}_{i}\rangle_{\text{Exact}}(\Vec{Q}_{\text{ML}}(t))$ is the ML prediction error at the $t$-th time step. $o(q(t))$ means higher order terms in $q(t)$ (See Supplemental Material for the derivation). Given a bounded $F(t)$, even if it is tuned to drive the oscillator optimally, as long as $K,\gamma > 0$, the oscillator cannot be driven to infinite amplitude, i.e. the accumulated momentum and position errors are bounded. For $F(t)$ to be bounded, it suffices to have $\delta\langle\hat{O}_i\rangle(t)$ bounded, which motivates us to define an Error Bounded Property such that when $\delta\langle\hat{O}_i\rangle(t)$ is bounded, $p(t)$ and $q(t)$ are also bounded. More precisely, we define the following
\begin{definition} [Error Bounded Property]
\label{def:error_bounded_property}
    A model satisfies the Error Bounded Property with respect to $\delta\langle\hat{O}_i\rangle(t)$ for $t\in[t_\text{init},t_\text{end}]$, if and only if the following claim is true: $\exists\, C_q, C_p>0$ such that $\forall \epsilon>0$, if $\forall t\in[t_\text{init},t_\text{end}]$, $|\delta\langle\hat{O}_i\rangle(t)|^2\le \epsilon$, then there are $|q(t)|\le C_q\sqrt{\epsilon}$ and $|p(t)|\le C_p\sqrt{\epsilon}$, $\forall t\in[t_\text{init},t_\text{end}]$.
\end{definition}
\begin{proposition}
\label{prop:approx_const_linear_model}
    The approximately constant linear model satisfies the Error Bounded Property if $K>0$.
\end{proposition}
The proof of Prop.~\ref{prop:approx_const_linear_model} is in Supplemental Material.

\textit{(ii) Non-linear model.} We can generalize the approximately constant linear model to allow non-linearity. We drop the assumption that the response $\frac{\partial\langle\hat{O}_{i}\rangle}{\partial Q_i}$ is approximately a constant, allow $G_{\alpha,i}(\Vec{P},\Vec{Q})$ to be a non-linear function of $Q_i$, and allow the potential $\frac{1}{2}k Q_i^2$ to include non-quadratic component in $Q_i$, as long as we can Taylor expand the potential at its minimum. These generalizations can be absorbed by a redefined $F(t)$ and a time-dependent error stiffness $K(t)$. Unlike the approximately constant linear model, the oscillator could have infinite amplitude even if $K(t)>0$ for all the time. We present a condition in Supplemental Material which guarantees a bounded bounded in the worst case scenario. Summarized as an informal theorem, we have:

\begin{figure*}[ht!]
 \centering
\includegraphics[width=2\columnwidth]{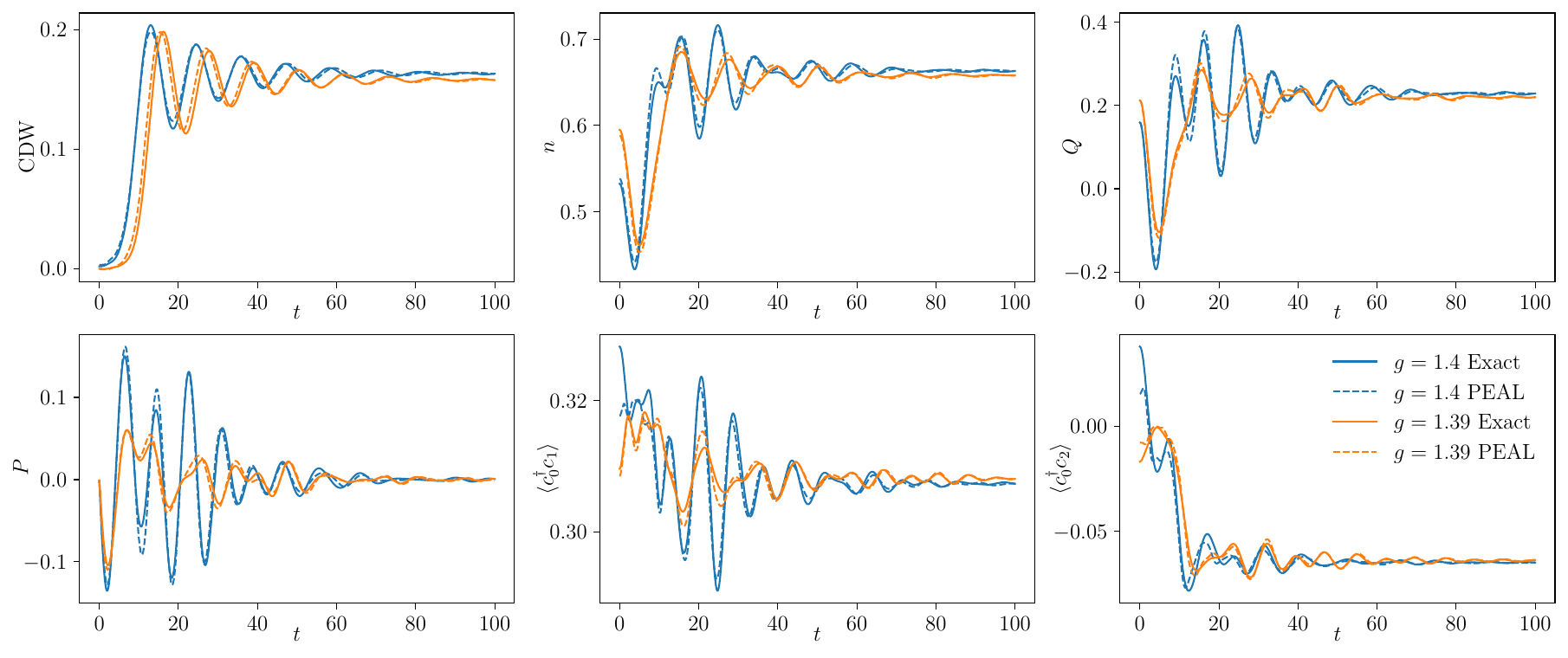}
\caption{PEAL (dashed) vs. exact simulation (solid) in single-path prediction. Blue curves correspond to standard learning with $g=1.4$, and orange curves correspond to transfer learning with $g=1.39$. }
\vspace{-15pt}
\label{fig:single_path}
\end{figure*} 

\begin{theorem} [Error Bounded Condition for non-linear model (Informal)]
\label{thm:error_convergence}
    If $K(t)>M(\gamma/2)^2+C$ with a positive constant $C$ for all $t$, and the error stiffness $K(t)$'s fluctuation, as well as $G_{\alpha,i}(\Vec{P},\Vec{Q})$ and its first derivatives are bounded, then the non-linear model satisfies the Error Bounded Property in Def.~\ref{def:error_bounded_property}.
\end{theorem}

The proof for Thm.~\ref{thm:error_convergence} is provided in Supplemental Material. We can further apply it to the Hamiltonian in normal mode with the quadratic momentum under Fourier transform, which could appear in a more general setup.

\textit{(iii) General Relaxation Method.} We now come back to the most general ADML. We allow arbitrary $G_{\alpha,i}(\Vec{P},\Vec{Q})$ and $H_{cl}(\Vec{P},\Vec{Q})$ in Eq.~(\ref{eq:ham}), and we make no assumption on the response $\frac{\partial\langle\hat{O}_{i}\rangle}{\partial Q_j}$. Because in the most general case the classical degrees of freedom are no longer decoupled, we restore the $i$ index of the accumulated momentum and position errors, $p_i(t) = P_{i,\text{ML}}(t)-P_{i,\text{Exact}}(t)$, $q_i(t) = Q_{i,\text{ML}}(t)-Q_{i,\text{Exact}}(t)$.

For a general ADML, the EOM for the errors $q_i(t)$, $p_i(t)$ can in general be derived from Eq.~(\ref{eq:eom_Q}) and~(\ref{eq:eom_P}):
\begin{align}
    \frac{d}{dt} q_i &= \sum_j \left( \mathcal{K}_{q_j,q_i} q_j + \mathcal{K}_{p_j,q_i} p_j \right) + \mathcal{F}_{q_i} + o(q,p),
    \label{eq:eom_q_i}
    \\
    \frac{d}{dt} p_i &= \sum_j \left( \mathcal{K}_{q_j,p_i} q_j + \mathcal{K}_{p_j,p_i} p_j \right) + \mathcal{F}_{p_i} - \gamma p_i + o(q,p),
    \label{eq:eom_p_i}
\end{align}
where the error stiffness matrix $\mathcal{K}_{q(p)_j,q(p)_i}$ and the error force vector $\mathcal{F}_{q(p)_i}$ depend on $(\Vec{P},\Vec{Q})$, and $\mathcal{F}_{q(p)_i}$ is linear in $\delta\langle\hat{O}_{\alpha, i}\rangle(t)$ (see Supplemental Material for details). 

While it is difficult to write down an error bounded condition for a general ADML, we propose a relaxation method to provide a sufficient (but not necessary) condition to check the Error Bounded Property in Def.~\ref{def:error_bounded_property}.
The idea is to consider the worst case scenario. If in the worst case scenario the error still converges, then it is safe to use ADML to accelerate our dynamical simulation. The relaxation method is as follows. First, we require $\mathcal{K}$ elements, as well as $G_{\alpha,i}(\Vec{P},\Vec{Q})$ and its first derivatives, are bounded. Second, assuming $|\delta\langle\hat{O}_{\alpha,i}\rangle(t)|^2\le \epsilon$, we identify possible upper and lower bounds for the elements in $\mathcal{K}$ and $\mathcal{F}$. 
The bounds just only need to be effective during the time range of the simulation. Third, to achieve the worst case scenario, we want to maximize $d q_i/dt$ when $q_i$ is positive, and minimize when negative (the same for $p_i$). Therefore, we insert the upper bound of $\mathcal{K}_{q(p)_j,q(p)_i}$ when $q(p)_i$ and $q(p)_j$ have the same sign, and the lower bound if the opposite sign. $\mathcal{F}_{q(p)_i}$ are adjusted to their upper or lower bounds accordingly. Fourth, we perform a classical simulation of the EOM in Eq.~(\ref{eq:eom_q_i}) and~(\ref{eq:eom_p_i}), with the worst case scenario stated above. Finally, if the simulation shows there exist constants $C_{q_i}$ and $C_{p_i}$ such that for any $\epsilon>0$, there are $|q_i(t)|\le C_{q_i}\sqrt{\epsilon}$ and $|p_i(t)|\le C_{p_i}\sqrt{\epsilon}$ during the time range of interest, then the Error Bounded Property is verified with the relaxation method.

\emph{Provably Efficient Adiabatic Learning.}--- Next, we present a provably efficient learning algorithm for the above ADML models based on quantum information theory, which we call \textit{Provable Efficient Adiabatic Learning} (PEAL). The PEAL algorithm, equipped with a learning model $\mathcal{M}$ and a classical ordinary differential equation (ODE) solver, consists of the following steps. 

(i) Data collection for training. We sample a set of $N_s$ initial conditions $\{(g_\alpha, \Vec{P}(t_{\text{init}}), \Vec{Q}(t_{\text{init}}))_s\}_{s=1}^{N_s}$ from a distribution $\mathcal{D}_{\text{init}}$. 
We evolve the system with QSS and ODE solver to get $(\Vec{P}(t), \Vec{Q}(t), \langle\hat{O}\rangle(t))_s$. 
For each $s$, we sample a set of time steps $\mathcal{T}_s$ uniformly from $[t_{\text{init}}, t_{\text{end}}]$. 
(ii) Model training. We use the dataset $\bigcup_{s=1}^{N_s}\{(g_\alpha, \Vec{P}(t), \Vec{Q}(t), \langle\hat{O}\rangle(t))_s : t\in\mathcal{T}_s\}$ to train a model $\mathcal{M}:(g_\alpha, \Vec{P}, \Vec{Q})\mapsto \langle\hat{O}\rangle$ with the learning algorithm developed in~\cite{Preskill:2024nature}. 
(iii) Prediction. For any unseen new initial condition $(g_\alpha, \Vec{P}(t_{\text{init}}), \Vec{Q}(t_{\text{init}}))_{\text{new}} \sim \mathcal{D}_{\text{init}}$, PEAL outputs the dynamical trajectory $(\Vec{P}(t), \Vec{Q}(t), \langle\hat{O}\rangle(t))_{\text{new}}$ for $t\in[t_{\text{init}}, t_{\text{end}}]$, by alternatively updating $\langle\hat{O}\rangle(t)$ with $\mathcal{M}$ and $\Vec{P}(t), \Vec{Q}(t)$ with ODE solver, integrated with our symmetry-preserving techniques shown later. The sample complexity and error bounds are summarized in the following informal theorem.

\begin{theorem} [Provably Efficient Adiabatic Learning Theorem (Informal)]
    \label{thm:PEAL}
   When the Error Bounded Property in Def.~\ref{def:error_bounded_property} is satisfied, for $T$ time steps quantum-classical adiabatic dynamics of an $n$-qubit gapped system, with sample complexity $O(\textup{log}(n))$, 
   PEAL gives rise to controllable accumulated errors of classical variables and all k-local, bounded quantum observables scaling as (i) $O(\sqrt{T})$ for generic model $\mathcal{M}$ (ii) $O(\sqrt{\log T})$ for sub-Gaussian $\mathcal{M}$'s prediction error (iii) independent on $T$ for bounded $\mathcal{M}$'s prediction error. 
\end{theorem}

We leave the proofs for the above theorem in Supplemental Material. We note that the computational time for PEAL's prediction under a fixed $g_{\alpha}$ is $O(\min\{n N_s, c(n)\}T)$, where $c(n)$ is the ODE solver per time step complexity and the $n N_s$ factor comes from the model $\mathcal{M}$~\cite{Preskill:2024nature}. For Theorem~\ref{thm:PEAL}, it can also be applied to an $n$-qubit system with ground states of exponential-decay correlation functions using recent results~\cite{onorati2023efficient}. We highlight that PEAL works for unseen couplings $g_{\alpha,\text{new}} \notin \{g_{\alpha,s}\}_{s=1}^{N_s}$, demonstrating transfer learning over a family of Hamiltonian. Furthermore, PEAL can predict any $k$-local observable $\hat{O}$ even if it does not appear in the Hamiltonian.

\emph{Numerical Experiments.}--- We consider the Holstein model~\cite{Holstein59,Noack91,Bonca99} which describes the electron phonon interaction as follows:
\begin{equation}
\label{eq:H_ep}
    H = - \sum_{i,j} c_i^{\dagger} c_j - g \sum_i \left(c_i^{\dagger}c_i - \frac{1}{2} \right) Q_i + \sum_{i} \left(\frac{P_i^2}{2M} + \frac{k Q_i^2}{2} \right),
\end{equation}
where $c_i$ is related to the fermionic degree of freedom, while $Q_i$ and $P_i$ are the position and momentum of the phonon as classical degrees of freedom. 

In the numerical experiment, we study such model on a 1D periodic chain with $L=50$ sites. We set $M=k=1$ and the damping coefficient $\gamma=0.1$. The electronic degree of freedom at time $t$ will be the ground state of $H_e(\{Q_i(t)\})$ due to the fast relaxation, showing a 1D strong Anderson localization~\cite{Anderson58}. Therefore, PEAL's requirement of a ground state with exponential-decay correlation functions is satisfied. Analytically, the 1D adiabatic Holstein model always cools down to a charge density wave (CDW) in zero temperature. In Supplemental Material, we derive the relation between phonon amplitude and CDW response function, showing the error stiffness $K$ is always positive when it's close to a checkerboard configuration. We also numerically measure the error stiffness $K(t)$ during the time range of interest, confirming the Error Bounded Property (see Supplemental Material). Therefore, we can apply PEAL to the 1D adiabatic Holstein model with a bounded error guarantee during the dynamics. 

\begin{figure}[b!]
\centering
\includegraphics[width=8.0cm]{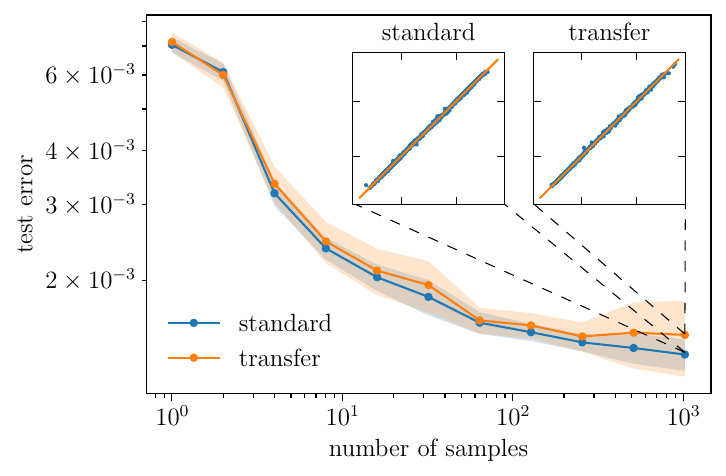}
\caption{Test errors for standard learning (blue) and transfer learning (orange). Inner pannels: $n_{\text{PEAL}}$ vs. $n_{\text{Exact}}$ for standard learning with $g \in G_\text{SL}$ (Left) and transfer learning with $g \in G_\text{TL}$ (Right). }
\label{fig:error_scaling}
\vspace{-20pt}
\end{figure}

Since Eq.~(\ref{eq:H_ep}) has the $U(1)$ and translation symmetry, we develop a symmetry-preserving PEAL. The $U(1)$ global symmetry is respected by conserving the total electron density, and the translation symmetry by applying the same model on all sites. The $U(1)$ symmetry preserving is a new feature in our PEAL that does not exist in previous literature and we have shown that the symmetry-preserving PEAL maintains provably efficient error bound in Supplemental Material.

\begin{figure}[t!]
\centering
\includegraphics[width=8.5cm]{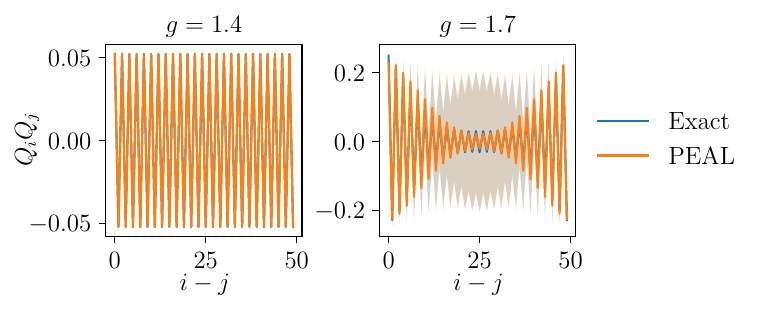}
\caption{Consistent agreement between PEAL and the exact simulation for the ensemble $Q_i(t)Q_j(t)$  correlation at $t=100$ of $g=1.4$ (Left) and $g=1.7$ (Right). } 
\label{fig:corr}
\vspace{-20pt}
\end{figure}

We begin with the single-path prediction task by training a model using some initial conditions and certain $g$ values. The goal is to predict observable dynamics from different initial conditions under both the training $g$ values (standard learning) and unseen $g$ values (transfer learning). Here, we choose $g\in G_\text{SL} = \{1.3, 1.32, 1.34, 1.36, 1.38, 1.4\}$ for training. In Figure~\ref{fig:single_path}, we demonstrate the single-path prediction by PEAL. The solid curves are the exact simulation using exact diagonalization (ED) for QSS and RK4 for classical ODE solver~\cite{RK4:textbook}, and the dashed curves are with PEAL. Blue curves are for $g=1.4$ (standard learning), and orange curves for $g=1.39$ (transfer learning). We present the time evolution of six different observables during the dynamics, which are the total charge density wave ($\text{CDW}) \sum_i(-1)^i n_i$, the electron density at the first site $n$, the phonon amplitude at the first site $Q$, the phonon momentum at the first site $P$, the hopping term $\langle c_0^\dagger c_1\rangle$, and the next-nearest-neighbor correlation $\langle c_0^\dagger c_2\rangle$. We note that CDW is a sum of local observables, $\langle c_0^\dagger c_1\rangle$ does not enter classical EOM, and $\langle c_0^\dagger c_2\rangle$ does not even appear in the Hamiltonian. Nevertheless, all observables in the PEAL curves agree well with their corresponding exact simulations, demonstrating that PEAL guarantees a controllable error for $k$-local observable and well performs in transfer learning. 

In Figure~\ref{fig:error_scaling}, we present the sample complexity of PEAL. We use the same training data in the single-path prediction, build the test set with $g\in G_\text{SL}$ but with different initial conditions, and demonstrate transfer learning with $g\in G_\text{TL} = \{1.31, 1.33, 1.35, 1.37, 1.39\}$. For illustration, we consider the root mean square of $|n_{\text{PEAL}}-n_{\text{Exact}}|$ as the test error. The test errors for both standard learning and transfer learning show a logarithmic scaling law with the number of samples. The inner panels also demonstrate that $n_{\text{PEAL}}$ in standard learning and transfer learning match well with $n_{\text{Exact}}$.

Generically, the ensemble dynamics (i.e.~the dynamics average over an ensemble of paths from different initial conditions) can provide more information of the underlying physics. In Figure~\ref{fig:corr}, we compare the ensemble averages of the correlation function $Q_i(t)Q_j(t)$ at time $t=100$ for $g=1.4$ and $g=1.7$. Here, $g=1.4$ prediction is based on PEAL training in $g \in G_{SL}$, and $g=1.7$ prediction is based on another model from PEAL training in $g\in\{1.6,1.62,1.64,1.66,1.68,1.7\}$ where long-lasting domain walls exist. For $g=1.4$, the system has cooled down to a pure checkerboard configuration, and shows an oscillating correlation function. For $g=1.7$, since the system has domain walls, it shows a decaying correlation function and a large variance. PEAL provides a good agreement with the exact simulations for both the mean and the variance over different $g$ values.

\emph{Conclusion.}--- In this work, we establish a comprehensive theoretical framework for analyzing quantum-classical adiabatic dynamics using learning algorithms. We provide a systematic analysis for the error bounded properties of the approximately constant linear model, the non-linear model, and generic models, ensuring the reliability of ADML. We develop a provably efficient adiabatic learning algorithm PEAL, demonstrating logarithmic scaling of sample complexity with system size and favorable evolution time scaling. Benchmarking PEAL on the Holstein model, we achieve accurate predictions of single-path dynamics and ensemble dynamics observables, with effective transfer learning across various quantum-classical coupling strengths. Our framework and algorithm opens up new directions for efficient learning in quantum-classical dynamics, including applying PEAL to quantum experiments, integrating advanced machine learning techniques, and extending to generic non-equilbrium processes.

\emph{Acknowledgements}---  The authors acknowledge helpful discussion with Max Metlitski, Lingyu Yang and Zhi Ren. DL acknowledges support from the NSF AI Institute for Artificial Intelligence and Fundamental Interactions (IAIFI). JPL acknowledges support by the National Science Foundation (PHY-1818914, CCF-1729369), the NSF Quantum Leap Challenge Institute (QLCI) program (OMA-2016245, OMA-2120757), and a Simons Foundation award (No. 825053). GWC acknowledges the support of the US Department of Energy Basic Energy Sciences under Contract No. DE-SC0020330.
The authors acknowledge the MIT SuperCloud for providing HPC resources that have contributed to the research results reported within this paper.

\bibliography{reference}
\clearpage
\onecolumngrid
\renewcommand\thefigure{S\arabic{figure}}  
\renewcommand\thetable{S\arabic{table}}  
\renewcommand{\theequation}{S\arabic{equation}}
\renewcommand{\thepage}{P\arabic{page}} 
\setcounter{page}{1}
\setcounter{figure}{0}  
\setcounter{table}{0}
\setcounter{equation}{0}

\def\beq{\begin{equation}}
\def\eeq{\end{equation}}
\appendix

\section{\label{app:lindblad} \large{Supplemental Material for Provably Efficient Machine Learning for Adiabatic Quantum-Classical Dynamics}}

\section{I. Dimensionless Model}

Consider the standard spinless Holstein model:
\begin{equation}
H = -t_{nn} \sum_{\langle ij \rangle} c_i^\dagger c_j - g \sum_i \left( c_i^\dagger c_i - \frac{1}{2} \right) Q_i + \sum_i \left( \frac{P_i^2}{2M} + \frac{kQ_i^2}{2} \right).
\end{equation}

And the equation of motion for the phonons is
\begin{equation}
\frac{dQ_i}{dt} = \frac{P_i}{M}, \quad \frac{dP_i}{dt} = g n_i - k Q_i,
\end{equation}
where $n_i = \langle c_i^\dagger c_i \rangle$ is the on-site fermion number. The mass $m$ and elastic constant $k$ are related by the familiar formula,
\begin{equation}
\omega = \sqrt{\frac{k}{M}},
\end{equation}
The inverse $\omega^{-1}$ gives a characteristic time scale for the dynamical problem. Next, one can introduce a ``length scale'' $Q_0$ for the displacement of the simple harmonic oscillator. The energy related to $Q$ at a given site is
\begin{equation}
E(Q) = -g n_i Q_i + \frac{k Q^2}{2}.
\end{equation}

Assuming electron number $n \sim 1$, minimization with respect to $Q$ gives $\partial E/\partial Q |_{Q_0} = 0$:
\begin{equation}
Q_0 \sim \frac{g}{k}.
\end{equation}

From this one can then introduce a scale for the momentum via the relation $dQ/dt = P/M$
\begin{equation}
\omega Q_0 \sim \frac{P_0}{M} \implies P_0 = M \omega Q_0 = \frac{M \omega g}{k}.
\end{equation}

We can now define the dimensionless time, displacement and momentum as
\begin{equation}
\tilde{t} = \omega t, \quad \tilde{Q}_i = \frac{Q_i}{Q_0}, \quad \tilde{P}_i = \frac{P_i}{P_0}.
\end{equation}

In terms of dimensionless quantities, the equation of motion is then simplified to
\begin{equation}
\frac{d\tilde{Q}_i}{d\tilde{t}} = \tilde{P}_i, \quad \frac{d\tilde{P}_i}{d\tilde{t}} = n_i - \tilde{Q}_i.
\end{equation}

Next, we consider the tight-binding Hamiltonian for the fermions for a given $\{Q_i\}$ configuration. We factor out the nearest-neighbor hopping constant $t_{nn}$ and use it as the unit for energy. Also, we use the dimensionless $\tilde{Q}_i$:
\begin{equation}
H_{TB} = t_{nn} \left[ -\sum_{\langle ij \rangle} c_i^\dagger c_j + \frac{g Q_0}{t_{nn}} \sum_i \tilde{Q}_i c_i^\dagger c_i \right].
\end{equation}

The coefficient of the second term above gives an important dimensionless parameter for Holstein model. Instead of $t_{nn}$, we can introduce the bandwidth of the tight-binding model: $W = 4t_{nn}$ for the 1D model. We then define a dimensionless electron-phonon coupling
\begin{equation}
\lambda = \frac{g Q_0}{W} = \frac{g^2}{kW}.
\end{equation}

The dimensionless tight-binding Hamiltonian then becomes
\begin{equation}
\tilde{H}_{TB} = -\sum_{\langle ij \rangle} c_i^\dagger c_j + 4\lambda \sum_i \tilde{Q}_i c_i^\dagger c_i.
\end{equation}

One can see that, using these dimensionless quantities, the only adjustable parameter of the adiabatic dynamics of the Holstein model is this dimensionless $\lambda$. In general, for real materials $\lambda \lesssim 1$. For example, we can set it to $\lambda = 0.5$ or $0.25$ in the simulations.

\section{II. Holstein model charge density wave response analysis}

Consider a tight binding model on a 1D lattice:
\begin{align}
    H &= -t \sum_{i} (c_i^{\dagger} c_{i+1} + h.c.) - g \sum_i (-1)^i Q c_i^{\dagger}c_i,
\end{align}
where $i\in\{0,1,\dots,L-1\}$, $L$ is the lattice size and is even. The lattice has periodic boundary condition. The effective potential on the lattice is staggered: $[-gQ, +gQ, -gQ, +gQ, \dots]$. The unit cell consists of 2 lattice sites. 

We can solve the single-particle wavefunctions using the ansatz
\begin{align}
    |\psi_k\rangle = (a, b e^{ik}, a e^{2ik}, b e^{3ik}, \dots)^\text{T}, \quad k\in \{0, \frac{2\pi}{L}, \dots, \frac{2\pi}{L}(\frac{L}{2}-1)\}.
\end{align}
Note that the range of $k$ is halved because the unit cell is doubled.

The eigenvalue equation $H|\psi_k\rangle = E|\psi_k\rangle$ becomes
\begin{align}
    - g Q a - 2(\cos k) b &= E a, \\
    - (2\cos k) a + g Q b &= E b,
\end{align}
which requires $E=\pm \sqrt{(gQ)^2+(2\cos k)^2}$ to have nontrivial solutions. 

The solutions form two bands. At half filling, all the states in the lower band, which has negative $E$, are filled with a particle. The charge density wave amplitude for $|\psi_k\rangle$ is $n_k=(a^2-b^2)/2$. From the eigenvalue equation and the normalization $(L/2)(a^2+b^2)=1$, we can solve the CDW amplitude for $|\psi_k\rangle$:
\begin{align}
    n_k = \frac{\text{sgn}(gQ)}{L\sqrt{(\frac{2\cos k}{g Q})^2 + 1}},
\end{align}
where $\text{sgn}(gQ)$ is the sign of $gQ$.

The total charge density wave at half filling is 
\begin{align}
    n = \sum_k n_k = \sum_k \frac{\text{sgn}(gQ)}{L\sqrt{(\frac{2\cos k}{g Q})^2 + 1}}.
\end{align}
The range of $k$ has been described above.

In the infinite lattice limit, $L\to\infty$, the sum over $k$ turns into an integral 
\begin{align}
    n \to \int_0^\pi \frac{dk}{2\pi} \frac{\text{sgn}(gQ)}{\sqrt{(\frac{2\cos k}{g Q})^2 + 1}} = \text{sgn}(gQ) \frac{\text{EllipticK}[-(\frac{2}{gQ})^2]}{\pi},
\label{eq:appendix_n_gQ}
\end{align}
where $\text{EllipticK}[m]\equiv \frac{\pi}{2}{}_2 F_1(\frac{1}{2}, \frac{1}{2};1;m)$ is the complete elliptic integral of the first kind, ${}_2 F_1(a, b;c;x)$ is the hypergeometric function~\cite{Mathematica:EllipticK}.

When $gQ\to 0$, the result says $n\to 0$. However, the derivative $\frac{\partial n}{\partial (gQ)}$ diverges at zero as $\log (\frac{1}{gQ})$. More precisely, $n\sim -gQ \log (gQ)$ for small $gQ$.

\begin{figure}[t]
\centering
\includegraphics[width=15cm]{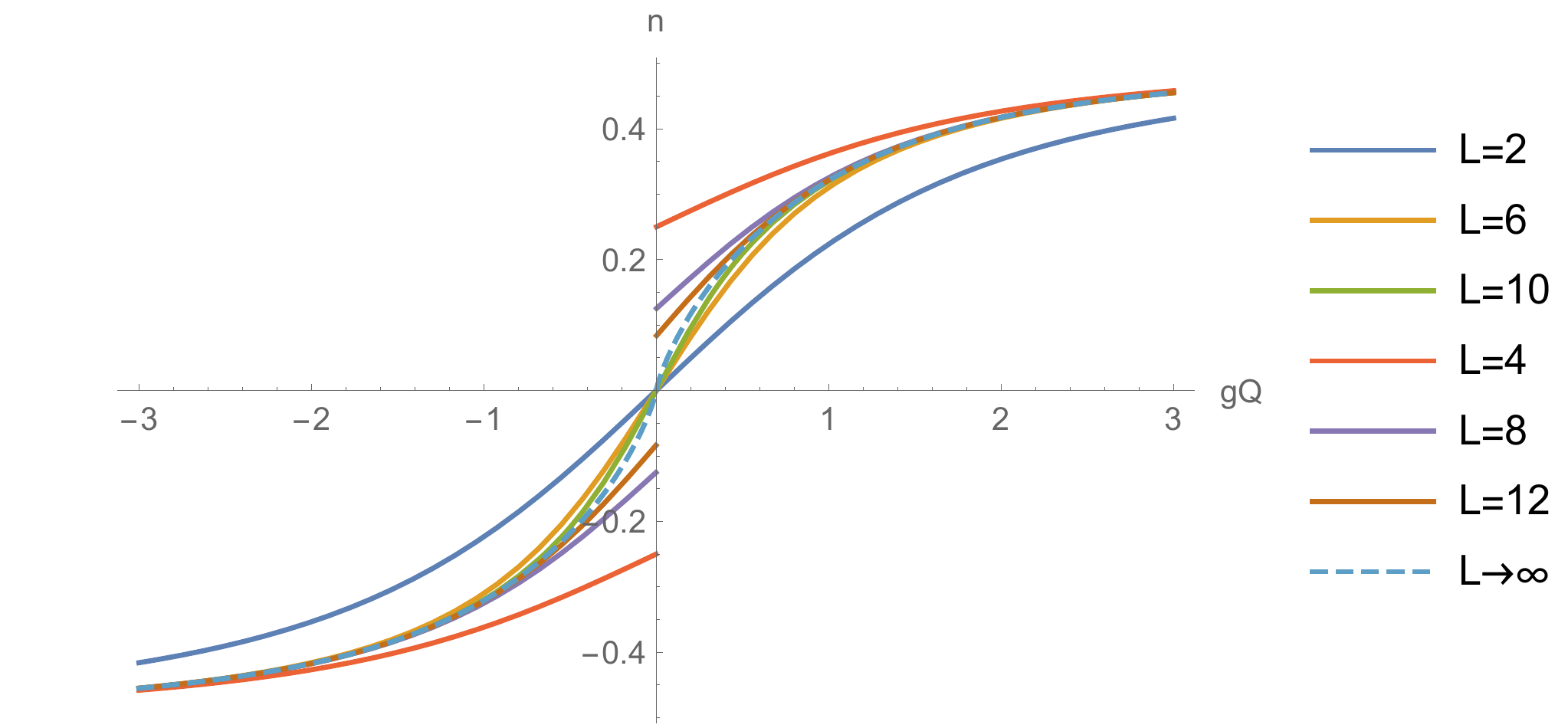}
\caption{The charge density wave amplitude $n$ as a function of the staggering potential $gQ$. For $L=4N$, the function has a discontinuity at zero: There is a constant $\text{CDW}=\frac{1}{L}$ even when $g\to 0$. For $L=4N+2$, the function is continuous and has a finite slope at zero: There is a critical slope for which a straight line cross the origin could have other crossing points with the function. i.e. There exists a critical $g$ value, below which the system does not have a stable CDW configuration. For $L\to\infty$ (dashed curve), the function is continuous and has a logarithmically diverging slope at zero: There is always CDW, but the amplitude is exponentially small for small $g$.}
\label{fig:rhon_Q}
\end{figure}

The property of $n$ as a function of $gQ$ is important, especially for $gQ$ close to zero. This is because when the system stabilizes, the equation of motion tells us $k_\text{spring}Q = g n$, i.e. the forces are balanced. This means a straight line $n=(k_\text{spring}/g^2)gQ$ that crosses the origin. The number of crossing points between this straight line and the $n$ vs. $gQ$ function determines whether the system have a stable CDW in the long time. Figure~\ref{fig:rhon_Q} shows the function curves for different lattice size.

For example, in the infinite lattice limit, the derivative diverges at zero. Therefore, any straight line crossing the origin with a finite slope will cross the function at some other points. This means that there always exists a stable CDW for the infinite lattice. However, we can estimate how large the CDW amplitude should be for a small $g$. Combine the straight line $n=(k_\text{spring}/g^2)gQ$ and the asymptotic behavior $n\sim -gQ \log (gQ)$, we can solve a non-trivial crossing point at the phonon amplitude $Q\sim e^{-k_\text{spring}/g^2}/g$. We see that the amplitude decays exponentially when $g\to0$.

However, for a finite system size $L$, the situation is different. For $L=4N$, the function $n$ has a finite limit $\pm\frac{1}{L}$ when $gQ\to0$, and has a discontinuity at $0$. Therefore, a straight line crossing the origin always has non-trivial intersections with the function curve. i.e. There exists CDW with amplitude at least $\frac{1}{L}$ for all values of $g$. For $L=4N+2$, the function $n$ goes to zero when $gQ\to0$, and has a finite slope at $0$. When $g$ is small, $(k_\text{spring}/g^2)$ exceeds this slope, the straight line only has intersection with the function curve at zero. When $g$ is large, $(k_\text{spring}/g^2)$ smaller than this slope, there are non-trivial intersections. Therefore, there is a phase transition due to the finite system size: When $g<g_{crit}$, no CDW; when $g>g_{crit}$, there is CDW. The system size we study, $L=50$, is in this situation.

We can analytically solve $g_{crit}$. We can take the derivative of Eq.~\ref{eq:appendix_n_gQ} at the limit to zero:
\begin{align}
    \left.\frac{\partial n}{\partial(gQ)}\right|_{gQ\to 0^{+}} = \sum_k \frac{1}{2L|\cos k|}.
\label{eq:appendix_slope}
\end{align}
By identifying $(k_\text{spring}/g_{crit}^2)$ with this slope, we get
\begin{align}
    g_{crit} = \sqrt{\frac{2k_\text{spring}L}{\sum_k \frac{1}{|\cos k|}}}.
\label{eq:appendix_g_crit}
\end{align}
Some numerical values for $k_\text{spring}=1$ are listed in Table.~\ref{table:g_crit}

\begin{table}[h!]
    \centering
    \begin{tabular}{|c|c|c|c|c|c|c|c|c|}
        \hline
        $L$ & $2$ & $6$ & $10$ & $22$ & $50$ & $102$ & $1002$ & $L \rightarrow \infty$ \\ \hline
        slope & $0.25$ & $0.4167$ & $0.4972$ & $0.6224$ & $0.7529$ & $0.8664$ & $1.2300$ & $\sim \log(L)$ \\ \hline
        $g_{crit}$ & $2$ & $1.5492$ & $1.4182$ & $1.2676$ & $1.1524$ & $1.0743$ & $0.9017$ & $\sim \frac{1}{\sqrt{\log L}}$ \\ \hline
    \end{tabular}
    \caption{Table of slope (Eq.~\ref{eq:appendix_slope}) and $g_{crit}$ values (Eq.~\ref{eq:appendix_g_crit}) for different $L$. We choose $k_\text{spring}=1$. }
    \label{table:g_crit}
\end{table}

For our choice $L=50$, $g_{crit}=1.1524$. The domain wall formation $g$ value for $L=50$ is around $1.6$. Therefore, we pick $g\in[1.3,1.4]$ in our numerical experiment, to avoid the finite size effect and the domain walls.

In the main text, we see the bounded error requires $k_\text{spring} - g\frac{\partial n}{\partial Q} > 0$. We can prove this is true when the system is close to the stable CDW configuration. We apply the self-consistency equation $k_\text{spring}=gn/Q$. From the concavity of the $n$ vs. $gQ$ function when $gQ>0$, and the convexity when $gQ<0$, we see $\frac{n}{gQ}>\frac{\partial n}{\partial (gQ)}$ as long as $gQ\neq 0$. Therefore, $k_\text{spring} - g\frac{\partial n}{\partial Q} = g^2 (\frac{n}{gQ} - \frac{\partial n}{\partial (gQ)}) > 0$ almost surely, because the $gQ=0$ case has measure $0$. This completes our proof.

\section{III. From the EOM of $Q$, $P$ to the EOM of $q$, $p$}

\subsection{Approximately constant linear model}

In the main text, we introduced the ``approximately constant linear model'', which has the Hamiltonian: 
\begin{align}
    \hat{H} = \hat{H}_{q} + \sum_{i} g \hat{O}_{i} Q_{i} + \sum_{i} \left( \frac{1}{2M} P_{i}^2 + \frac{1}{2} k Q_{i}^2 \right).
\end{align}
It has the EOM for $Q_i$ and $P_i$:
\begin{align}
    \frac{d}{dt} Q_i(t) &= \frac{1}{M} P_i(t), \\
    \frac{d}{dt} P_i(t) &= - g \langle\hat{O}_{i}\rangle -k Q_i(t) - \gamma P_i(t),
\end{align}
where the quantum observable $\langle\hat{O}_{i}\rangle$ is computed based on the configuration $\Vec{Q}(t)$.

The EOM above is equally valid for both the exact dynamical path and the ML dynamical path:
\begin{align}
    \frac{d}{dt} Q_{i,\text{Exact}}(t) &= \frac{1}{M} P_{i,\text{Exact}}(t), \\
    \frac{d}{dt} P_{i,\text{Exact}}(t) &= - g \langle\hat{O}_{i}\rangle_{\text{Exact}} -k Q_{i,\text{Exact}}(t) - \gamma P_{i,\text{Exact}}(t),
\end{align}
and 
\begin{align}
    \frac{d}{dt} Q_{i,\text{ML}}(t) &= \frac{1}{M} P_{i,\text{ML}}(t), \\
    \frac{d}{dt} P_{i,\text{ML}}(t) &= - g \langle\hat{O}_{i}\rangle_{\text{ML}} -k Q_{i,\text{ML}}(t) - \gamma P_{i,\text{ML}}(t).
\end{align}

Take the difference between the ML EOM and the exact EOM, define the accumulated position and momentum errors $q(t) = Q_{i,\text{ML}}(t)-Q_{i,\text{Exact}}(t)$, $p(t) = P_{i,\text{ML}}(t)-P_{i,\text{Exact}}(t)$, we get 
\begin{align}
    \frac{d}{dt} q(t) &= \frac{1}{M} p(t), \\
    \frac{d}{dt} p(t) &= - g (\langle\hat{O}_{i}\rangle_{\text{ML}} - \langle\hat{O}_{i}\rangle_{\text{Exact}}) -k q(t) - \gamma p(t).
\end{align}

However, here the observables depend on different configurations, $\langle\hat{O}_{i}\rangle_{\text{ML}} = \langle\hat{O}_{i}\rangle_{\text{ML}}(\Vec{Q}_{\text{ML}}(t))$, and $\langle\hat{O}_{i}\rangle_{\text{Exact}} = \langle\hat{O}_{i}\rangle_{\text{Exact}}(\Vec{Q}_{\text{Exact}}(t))$. When we take their difference, we not only need to take care about the difference between the prediction methods, we also need to take care about the difference between the configurations $\Vec{Q}_{\text{ML}}$ and $\Vec{Q}_{\text{Exact}}$. 

We do a Taylor expansion and apply the assumption that the off-diagonal response $\frac{\partial\langle\hat{O}_{i}\rangle}{\partial Q_j}$ ($i\neq j$) is approximately zero. We can make the difference between the observables into two terms:
\begin{align}
    \langle\hat{O}_{i}\rangle_{\text{ML}}(\Vec{Q}_{\text{ML}}(t)) - \langle\hat{O}_{i}\rangle_{\text{Exact}}(\Vec{Q}_{\text{Exact}}(t)) &= \langle\hat{O}_{i}\rangle_{\text{ML}}(\Vec{Q}_{\text{ML}}(t)) - \langle\hat{O}_{i}\rangle_{\text{Exact}}(\Vec{Q}_{\text{ML}}(t)) + \frac{\partial\langle\hat{O}_{i}\rangle}{\partial Q_i} q(t) + o(q(t)) 
    \nonumber \\
    &= \delta\langle\hat{O}_i\rangle(t) + \frac{\partial\langle\hat{O}_{i}\rangle}{\partial Q_i} q(t) + o(q(t)), 
\end{align}
where $\delta\langle\hat{O}_i\rangle(t) = \langle\hat{O}_{i}\rangle_{\text{ML}}(\Vec{Q}_{\text{ML}}(t)) - \langle\hat{O}_{i}\rangle_{\text{Exact}}(\Vec{Q}_{\text{ML}}(t))$ is the single-step prediction error.

Insert the two terms back to the EOM, we get the same result as in the main text:
\begin{align}
    \frac{d}{dt} q(t) &= \frac{1}{M} p(t),
    \\
    \frac{d}{dt} p(t) &= - g \delta\langle\hat{O}_i\rangle(t) - g \frac{\partial\langle\hat{O}_i\rangle}{\partial Q_i} q(t) - k q(t) - \gamma p(t) + o(q(t)).
\end{align}

\subsection{Non-linear model}

In the main text, we then generalized a bit to the non-linear model, which has the Hamiltonian: 
\begin{align}
    \hat{H} = \hat{H}_{q} + \sum_{i} g \hat{O}_{i} G_i(Q_{i}) + \sum_{i} \left( \frac{1}{2M} P_{i}^2 + V_i(Q_{i}) \right).
\end{align}
It has the EOM for $Q_i$ and $P_i$:
\begin{align}
    \frac{d}{dt} Q_i(t) &= \frac{1}{M} P_i(t), \\
    \frac{d}{dt} P_i(t) &= - g \langle\hat{O}_{i}\rangle G_i^{'}(Q_{i}) - V_i^{'}(Q_{i}) - \gamma P_i(t),
\end{align}
where the quantum observable $\langle\hat{O}_{i}\rangle$ is still computed based on the configuration $\Vec{Q}(t)$.

The EOM above is equally valid for both the exact dynamical path and the ML dynamical path:
\begin{align}
    \frac{d}{dt} Q_{i,\text{Exact}}(t) &= \frac{1}{M} P_{i,\text{Exact}}(t), \\
    \frac{d}{dt} P_{i,\text{Exact}}(t) &= - g \langle\hat{O}_{i}\rangle_{\text{Exact}} G_i^{'}(Q_{i,\text{Exact}}) -V_i^{'}(Q_{i,\text{Exact}}) - \gamma P_{i,\text{Exact}}(t),
\end{align}
and 
\begin{align}
    \frac{d}{dt} Q_{i,\text{ML}}(t) &= \frac{1}{M} P_{i,\text{ML}}(t), \\
    \frac{d}{dt} P_{i,\text{ML}}(t) &= - g \langle\hat{O}_{i}\rangle_{\text{ML}} G_i^{'}(Q_{i,\text{ML}}) -V_i^{'}(Q_{i,\text{ML}}) - \gamma P_{i,\text{ML}}(t).
\end{align}

Take the difference between the ML EOM and the exact EOM, define the accumulated position and momentum errors $q(t) = Q_{i,\text{ML}}(t)-Q_{i,\text{Exact}}(t)$, $p(t) = P_{i,\text{ML}}(t)-P_{i,\text{Exact}}(t)$, expand to first order of $q(t)$, we get:
\begin{align}
    \frac{d}{dt} q(t) &= \frac{1}{M} p(t), \\
    \frac{d}{dt} p(t) &= - g (\langle\hat{O}_{i}\rangle_{\text{ML}} - \langle\hat{O}_{i}\rangle_{\text{Exact}}) G_i^{'}(Q_{i,\text{ML}}) - g \langle\hat{O}_{i}\rangle_{\text{Exact}} G_i^{''}(Q_{i,\text{Exact}}) q(t) - V_i^{''}(Q_{i,\text{Exact}}) q(t) - \gamma p(t) + o(q(t)).
\end{align}

We apply the same technique as in the approximately constant linear model. Insert
\begin{align}
    \langle\hat{O}_{i}\rangle_{\text{ML}} - \langle\hat{O}_{i}\rangle_{\text{Exact}} = \delta\langle\hat{O}_i\rangle(t) + \frac{\partial\langle\hat{O}_{i}\rangle}{\partial Q_i} q(t) + o(q(t)), 
\end{align}
we get
\begin{align}
    \frac{d}{dt} q(t) &= \frac{1}{M} p(t), \\
    \frac{d}{dt} p(t) &= - g \delta\langle\hat{O}_i\rangle(t) G_i^{'}(Q_{i,\text{ML}}) - g \frac{\partial\langle\hat{O}_{i}\rangle}{\partial Q_i} G_i^{'}(Q_{i,\text{ML}}) q(t) - g \langle\hat{O}_{i}\rangle_{\text{Exact}} G_i^{''}(Q_{i,\text{Exact}}) q(t) - V_i^{''}(Q_{i,\text{Exact}}) q(t) - \gamma p(t) + o(q(t)).
    \nonumber \\
    &= F(t) - K(t) q(t) - \gamma p(t) + o(q(t)),
\end{align}
where we define the effective driving force and the effective spring constant as
\begin{align}
\label{eq:appendix_F_nonlinear}
    F(t) &= -g \delta\langle\hat{O}_i\rangle(t) G_i^{'}(Q_{i,\text{ML}}(t)), \\
    K(t) &= g \frac{\partial\langle\hat{O}_{i}\rangle}{\partial Q_i} G_i^{'}(Q_{i,\text{ML}}(t)) + g \langle\hat{O}_{i}\rangle_{\text{Exact}} G_i^{''}(Q_{i,\text{Exact}}(t)) + V_i^{''}(Q_{i,\text{Exact}}(t)).
\end{align}

This EOM is equivalent to a damped spring with time dependent driving force and spring constant. We can prove the amplitude of the spring will not diverge if $F(t)$ is bounded and $K(t)$ does not fluctuate largely.

\subsection{Generic model}

The generic model has the Hamiltonian 
\begin{align}
    \hat{H} = \hat{H}_{q} + \sum_{\alpha,i} g_\alpha \hat{O}_{\alpha,i} G_{\alpha,i}(\Vec{P},\Vec{Q}) + H_{cl}(\Vec{P},\Vec{Q}),
\end{align}
and the EOM
\begin{align}
    \frac{d}{dt} Q_j &= \sum_{\alpha,i} g_\alpha \langle\hat{O}_{\alpha,i}\rangle \frac{\partial}{\partial P_j} G_{\alpha,i} + \frac{\partial}{\partial P_j} H_{cl}, 
    \\
    \frac{d}{dt} P_j &= -\sum_{\alpha,i} g_\alpha \langle\hat{O}_{\alpha,i}\rangle \frac{\partial}{\partial Q_j} G_{\alpha,i} - \frac{\partial}{\partial Q_j} H_{cl} - \gamma P_j.
\end{align}

We can repeat the techniques above: writing down the EOM for ML and the exact simulation, making a difference of the EOM, defining $q_i$ and $p_i$ (here the index $i$ cannot be suppressed), doing Taylor expansion, and combining terms up to the first order. After similar derivations, we can get the result
\begin{align}
    \frac{d}{dt} q_i &= \sum_j \left[ q_j \mathcal{K}_{q_j,q_i}(\Vec{P},\Vec{Q}) + p_j \mathcal{K}_{p_j,q_i}(\Vec{P},\Vec{Q}) \right] + \mathcal{F}_{q_i} + o(q,p),
    \\
    \frac{d}{dt} p_i &= \sum_j \left[ q_j \mathcal{K}_{q_j,p_i}(\Vec{P},\Vec{Q}) + p_j \mathcal{K}_{p_j,p_i}(\Vec{P},\Vec{Q}) \right] + \mathcal{F}_{p_i} - \gamma p_i + o(q,p),
\end{align}
where
\begin{align}
    \mathcal{K}_{q_j,q_i} &= \sum_{\alpha,k} g_\alpha \left( \frac{\partial \langle\hat{O}_{\alpha,k}\rangle}{\partial Q_j} \frac{\partial G_{\alpha,k}}{\partial P_i} + \langle\hat{O}_{\alpha,k}\rangle \frac{\partial^2 G_{\alpha,k}}{\partial Q_j \partial P_i} \right) + \frac{\partial^2 H_{cl}}{\partial Q_j \partial P_i},
    \\
    \mathcal{K}_{p_j,q_i} &= \sum_{\alpha,k} g_\alpha \left( \frac{\partial \langle\hat{O}_{\alpha,k}\rangle}{\partial P_j} \frac{\partial G_{\alpha,k}}{\partial P_i} + \langle\hat{O}_{\alpha,k}\rangle \frac{\partial^2 G_{\alpha,k}}{\partial P_j \partial P_i} \right) + \frac{\partial^2 H_{cl}}{\partial P_j \partial P_i},
    \\
    \mathcal{K}_{q_j,p_i} &= - \sum_{\alpha,k} g_\alpha \left( \frac{\partial \langle\hat{O}_{\alpha,k}\rangle}{\partial Q_j} \frac{\partial G_{\alpha,k}}{\partial Q_i} + \langle\hat{O}_{\alpha,k}\rangle \frac{\partial^2 G_{\alpha,k}}{\partial Q_j \partial Q_i} \right) - \frac{\partial^2 H_{cl}}{\partial Q_j \partial Q_i},
    \\
    \mathcal{K}_{p_j,p_i} &= - \sum_{\alpha,k} g_\alpha \left( \frac{\partial \langle\hat{O}_{\alpha,k}\rangle}{\partial P_j} \frac{\partial G_{\alpha,k}}{\partial Q_i} + \langle\hat{O}_{\alpha,k}\rangle \frac{\partial^2 G_{\alpha,k}}{\partial P_j \partial Q_i} \right) - \frac{\partial^2 H_{cl}}{\partial P_j \partial Q_i},
    \\
    \mathcal{F}_{q_i} &= \sum_{\alpha,j} g_\alpha \delta\langle\hat{O}_{\alpha,j}\rangle(t)\frac{\partial}{\partial P_i}G_{\alpha,j}(\Vec{P},\Vec{Q}),
    \\
    \mathcal{F}_{p_i} &= -\sum_{\alpha,j} g_\alpha \delta\langle\hat{O}_{\alpha,j}\rangle(t)\frac{\partial}{\partial Q_i}G_{\alpha,j}(\Vec{P},\Vec{Q}).
\end{align}

\section{IV. Proof of Proposition~\ref{prop:approx_const_linear_model}}

Here we prove that the approximately constant linear model satisfies the Error Bounded Property in Def.~\ref{def:error_bounded_property} if $K>0$. We use the analogy to a damped harmonic oscillating spring, phrasing the error force $F(t)$ in the main text as the ``driving force'' and the error stiffness $K$ in the main text as the ``spring constant''. For a damped spring, given a bounded driving force, even if the force is tuned to drive the spring optimally, as long as $K, \gamma>0$, the spring cannot be driven to infinite amplitude. This can be seen by the following worst-case analysis. In a worst case scenario, the driving force is set to be the maximal value in the direction of the spring movement. We know a constant driving force means a shift of the reference point of the spring. Therefore, by shifting the reference point back and forth, the driving force at most linearly increases the amplitude with the motion cycles of the spring. However, a finite damping decreases the amplitude of the spring by a constant factor in each cycle. Therefore, for a large enough initial amplitude, the decrement of the amplitude due to the damping must exceed the increment of the amplitude due to the driving force in the cycle, and thus the amplitude in the next cycle must be smaller than the initial one, which tells us that the damped spring cannot be driven to infinite amplitude.

Now we analyze the relation between the maximal amplitude $\Bar{q}$, the maximal momentum $\Bar{p}$, and the maximal driving force $\Bar{F}$, using the dimensional analysis. At $t_\text{init}$, there are no accumulated errors, i.e. $q(t_\text{init})=p(t_\text{init})=0$. Therefore, the only physical quantity that carries the dimension of length is the maximal driving force $[\Bar{F}]=[M][L][T]^{-2}$, where $[M]$, $[L]$, and $[T]$ are the dimension of mass, length, and time, respectively. Note that the maximal amplitude $[\Bar{q}]=[L]$ and the maximal momentum $\Bar{p}=[L][T]^{-1}$ both carries power one of the length dimension $[L]$. Therefore, by dimensional analysis, they are both proportional to the maximal driving force $\Bar{F}$. i.e. $\exists\ \Bar{C}_q, \Bar{C}_p$ such that $\Bar{q} = \Bar{C}_q \Bar{F}$, $\Bar{p} = \Bar{C}_p \Bar{F}$.

For any $\epsilon>0$, if $|\delta\langle\hat{O}_i\rangle(t)|^2\le\epsilon$, by the definition $F(t)=-g\delta\langle\hat{O}_i\rangle(t)$, we have $|F(t)|\le |g|\sqrt{\epsilon}$. By the analogy above, it means that the maximal force $\Bar{F}=|g|\sqrt{\epsilon}$. Therefore, by the meaning of the maximal amplitude and the maximal momentum, there are $|q(t)|\le \Bar{q}=\Bar{C}_q \Bar{F}=\Bar{C}_q|g|\sqrt{\epsilon}$ and $|p(t)|\le \Bar{p}=\Bar{C}_p \Bar{F}=\Bar{C}_p|g|\sqrt{\epsilon}$. Define two new constants $C_q = |g|\Bar{C}_q$ and $C_p = |g|\Bar{C}_p$, we get $|q(t)|\le C_q\sqrt{\epsilon}$ and $|p(t)|\le C_p\sqrt{\epsilon}$, which is what we want to show. This completes the proof of Prop.~\ref{prop:approx_const_linear_model}. \qed

\section{V. Proof of Theorem~\ref{thm:error_convergence} (Error Bounded Condition for non-linear model)}

Here we present the formal version of Thm.~\ref{thm:error_convergence} in the main text. We use the analogy to a damped harmonic oscillating spring, phrasing the error force $F(t)$ in the main text as the ``driving force'' and the error stiffness $K(t)$ in the main text as the ``spring constant''.

\begin{theorem}[Error Bounded Condition for non-linear model]
\label{thm:appendix_error_convergence}

    Consider a damped harmonic oscillator with the following EOM:
    \begin{align}
        \frac{d}{dt} q(t) &= \frac{1}{M} p(t), \\
        \frac{d}{dt} p(t) &= F(t) - K(t) q(t) - \gamma p(t) + o(q(t)),
    \end{align}
    where $q$ is the classical coordinate, $p$ is the classical momentum, $t$ is time, $M$ is the mass of the oscillator, $F(t)$ is a time-dependent driving force, $K(t)$ is a time-dependent spring constant, $\gamma$ is the damping coefficient.The initial condition is $q(t_\text{init})=p(t_\text{init})=0$.

    Given $K_{\max}>K_{\min}>M(\gamma/2)^2$ satisfying the following inequality:
    \begin{align}
        \frac{K_{\max}}{K_{\min}} < \exp\left[2\left(\frac{\text{arctan}\ \omega_{\min}}{\omega_{\min}} + \frac{\pi - \text{arctan}\ \omega_{\max}}{\omega_{\max}}\right)\right],
        \label{eq:error_convergence_condition}
    \end{align}
    where $\omega_{\max}=\sqrt{\frac{K_{\max}}{M(\gamma/2)^2}-1}$, $\omega_{\min}=\sqrt{\frac{K_{\min}}{M(\gamma/2)^2}-1}$.

    If $K(t)\in[K_{\min},K_{\max}]$ for all $t$, and $G_{\alpha,i}(\Vec{P},\Vec{Q})$ and its first derivatives are bounded, then the Error Bounded Property in Def.~\ref{def:error_bounded_property} is satisfied, i.e. $\exists\ C_q, C_p>0$ such that $\forall \epsilon>0$, if $|\delta\langle\hat{O}_i\rangle(t)|^2\le \epsilon$ for all $t$, then there are 
    \begin{align}
        |q(t)| \le C_q\sqrt{\epsilon}, \quad |p(t)| \le C_p\sqrt{\epsilon}, \quad \text{for all } t.
    \end{align}
\end{theorem}

\textbf{Proof.} Consider a damped harmonic oscillator with the following EOM:
\begin{align}
    \frac{d}{dt} q(t) &= \frac{1}{M} p(t), \\
    \frac{d}{dt} p(t) &= F(t) - K(t) q(t) - \gamma p(t) + o(q(t)),
\end{align}
where $q$ is the classical coordinate, $p$ is the classical momentum, $t$ is time, $M$ is the mass of the oscillator, $F(t)$ is a time-dependent driving force, $K(t)$ is a time-dependent spring constant, $\gamma > 0$ is the damping coefficient. We assume $F(t)\in[-\Bar{F},\Bar{F}]$, $K(t)\in[K_{\min}, K_{\max}]$, where $K_{\max} > K_{\min} > M(\gamma/2)^2$. Note that with Eq.~\ref{eq:appendix_F_nonlinear}, $F(t) = -g \delta\langle\hat{O}_i\rangle(t) G_i^{'}(Q_{i,\text{ML}}(t))$, the assumption $|\delta\langle\hat{O}_i\rangle(t)|^2\le \epsilon$ implies $\Bar{F}\le O(\sqrt{\epsilon})$ given bounded $G_i^{'}$.

We consider the worst case scenario. The mass on the spring starts from one side with a large amplitude. As the mass moving towards the other side, the driving force always keeps the maximal value towards the other side. Before the mass crosses the origin, the spring constant is set as $K_{\max}$ to maximize the drag towards the other side. After the mass crosses the origin, the spring constant is set as $K_{\min}$ to minimize the burden for its moving as far as possible. In this worst case scenario, the mass will stop at its largest amplitude on the other side. If this amplitude on the other side is smaller than the one it started with, than the spring cannot have diverging amplitude, and therefore the error converges.

We now analyze this worst case scenario. Assume the mass starts at $q(-t_0)=-q_0$, $\Dot{q}(-t_0)=0$. At time $t=0$ it crosses the origin, $q(0)=0$, with a velocity $\Dot{q} (0) = v$. At time $t=t_1$ it stops on the other side at $q(t_1)=q_1$, $\Dot{q}(t_1)=0$.

Before the mass crosses the origin, the spring constant is $K_{\max}$. The EOM is
\begin{align}
    \Ddot{q}(t) = - \gamma \Dot{q}(t) -\frac{K_{\max}}{M} q(t) + \frac{\Bar{F}}{M}, \quad t\in[-t_0, 0].
\end{align}
The EOM has a general solution
\begin{align}
    q(t) = A e^{-\frac{\gamma}{2} t} \sin \Omega_{\max} t + B e^{-\frac{\gamma}{2} t} \cos \Omega_{\max} t + \frac{\Bar{F}}{K_{\max}},
\end{align}
where $\Omega_{\max} = \sqrt{\frac{K_{\max}}{M} - (\frac{\gamma}{2})^2}$.

Inserting the boundary condition at $t=0$, we get equations
\begin{align}
    q(0) &= 0 = B + \frac{\Bar{F}}{K_{\max}}, \\
    \Dot{q}(0) &= v = \Omega_{\max} A - \frac{\gamma}{2} B,
\end{align}
from which we solve
\begin{align}
    A &= \frac{1}{\Omega_{\max}}\left(v-\frac{\gamma}{2}\frac{\Bar{F}}{K_{\max}}\right), \\
    B &= -\frac{\Bar{F}}{K_{\max}}.
\end{align}

Inserting the boundary condition at $t=-t_0$, we get equations
\begin{align}
    q(-t_0) &= -q_0 = - A e^{\frac{\gamma}{2} t_0} \sin \Omega_{\max} t_0 + B e^{\frac{\gamma}{2} t_0} \cos \Omega_{\max} t_0 + \frac{\Bar{F}}{K_{\max}}, \\
    \Dot{q}(-t_0) &= 0 = A \left(\frac{\gamma}{2} e^{\frac{\gamma}{2} t_0} \sin \Omega_{\max} t_0 + \Omega_{\max} e^{\frac{\gamma}{2} t_0} \cos \Omega_{\max} t_0 \right) + B\left( - \frac{\gamma}{2} e^{\frac{\gamma}{2} t_0} \cos \Omega_{\max} t_0 + \Omega_{\max} e^{\frac{\gamma}{2} t_0} \sin \Omega_{\max} t_0 \right).
\end{align}
We can solve $t_0$ from the second equation:
\begin{align}
    t_0 &= \frac{1}{\Omega_{\max}} \text{arccot}\frac{(\gamma/2)A+\Omega_{\max}B}{-\Omega_{\max}A+(\gamma/2)B}
    \nonumber \\
    &= \frac{1}{\Omega_{\max}} \text{arccot}\left(-\frac{(\gamma/2)v - \Bar{F}/M}{\Omega_{\max}v}\right),
\end{align}
where we have inserted the solution of $A$ and $B$. We can then insert everything into the first equation to solve $q_0$, but we do not do it now.

Let's also take a look at the time after the mass crosses the origin. The spring constant changes to $K_{\min}$. The EOM is 
\begin{align}
    \Ddot{q}(t) = - \gamma \Dot{q}(t) -\frac{K_{\min}}{M} q(t) + \frac{\Bar{F}}{M}, \quad t\in[0, t_1].
\end{align}
The EOM has a general solution
\begin{align}
    q(t) = C e^{-\frac{\gamma}{2} t} \sin \Omega_{\min} t + D e^{-\frac{\gamma}{2} t} \cos \Omega_{\min} t + \frac{\Bar{F}}{K_{\min}},
\end{align}
where $\Omega_{\min} = \sqrt{\frac{K_{\min}}{M} - (\frac{\gamma}{2})^2}$.

Inserting the boundary condition at $t=0$, we get equations
\begin{align}
    q(0) &= 0 = D + \frac{\Bar{F}}{K_{\min}}, \\
    \Dot{q}(0) &= v = \Omega_{\min} C - \frac{\gamma}{2} D,
\end{align}
from which we solve
\begin{align}
    C &= \frac{1}{\Omega_{\min}}\left(v-\frac{\gamma}{2}\frac{\Bar{F}}{K_{\min}}\right), \\
    D &= -\frac{\Bar{F}}{K_{\min}}.
\end{align}

Inserting the boundary condition at $t=t_1$, we get equations
\begin{align}
    q(t_1) &= q_1 = C e^{-\frac{\gamma}{2} t_1} \sin \Omega_{\min} t_1 + D e^{-\frac{\gamma}{2} t_1} \cos \Omega_{\min} t_1 + \frac{\Bar{F}}{K_{\min}}, \\
    \Dot{q}(t_1) &= 0 = C \left( - \frac{\gamma}{2} e^{-\frac{\gamma}{2} t_1} \sin \Omega_{\min} t_1 + \Omega_{\min} e^{-\frac{\gamma}{2} t_1} \cos \Omega_{\min} t_1 \right) + D\left( - \frac{\gamma}{2} e^{-\frac{\gamma}{2} t_1} \cos \Omega_{\min} t_1 - \Omega_{\min} e^{-\frac{\gamma}{2} t_1} \sin \Omega_{\min} t_1 \right).
\end{align}
We can solve $t_1$ from the second equation:
\begin{align}
    t_1 &= \frac{1}{\Omega_{\min}} \text{arccot}\frac{(\gamma/2)C+\Omega_{\min}D}{\Omega_{\min}C-(\gamma/2)D}
    \nonumber \\
    &= \frac{1}{\Omega_{\min}} \text{arccot}\frac{(\gamma/2)v - \Bar{F}/M}{\Omega_{\min}v},
\end{align}
where we have inserted the solution of $C$ and $D$. Note that comparing to the case before, there is no minus sign in the arccot function. 

To have a bounded error, we want that for large enough $q_0$, there is $\frac{q_1}{q_0}<1$, where $q_0$ is the starting amplitude and $q_1$ is the stopping amplitude. i.e. The amplitude never diverges. Iteratively we consider $q_1$ as the next starting amplitude, and we can see the sequence of amplitudes converges in the long time. Note that the amplitude has the dimension of length, and among all given parameters, only the maximal force $\Bar{F}$ carries the dimension of length. Using dimension analysis, we see that the maximal amplitude the system could reach will be proportional to the maximal force $\Bar{F}$. The maximal momentum the system could reach will also be proportional to $\Bar{F}$ from dimension analysis. 
i.e. $q(t) \le O(\Bar{F})$, $p(t) \le O(\Bar{F})$. Combining with the assumption $\Bar{F} \le O(\sqrt{\epsilon})$, it leads to $q(t) \le O(\sqrt{\epsilon})$, $p(t) \le O(\sqrt{\epsilon})$, which is what we want to show.

Now the final step is to simplify the condition $\frac{q_1}{q_0}<1$. Note that for arbitrarily large $q_0$, the velocity $v$ can also be arbitrarily large. Therefore, we can consider the limit $\Bar{F}/v\to 0$. Under this limit,
\begin{align}
    A/v &\to \frac{1}{\Omega_{\max}}, \\
    B/v &\to 0, \\
    C/v &\to \frac{1}{\Omega_{\min}}, \\
    D/v &\to 0, \\
    t_0 &\to \frac{1}{\Omega_{\max}} \text{arccot}\left(-\frac{(\gamma/2)}{\Omega_{\max}}\right) = \frac{\pi - \text{arctan}(2\Omega_{\max}/\gamma)}{\Omega_{\max}}, \label{eq:appendix_t_0} \\
    t_1 &\to \frac{1}{\Omega_{\min}} \text{arccot}\frac{(\gamma/2)}{\Omega_{\min}} = \frac{\text{arctan}(2\Omega_{\min}/\gamma)}{\Omega_{\min}}. \label{eq:appendix_t_1}
\end{align}

Finally,
\begin{align}
    \frac{q_1}{q_0} &= \frac{C e^{-\frac{\gamma}{2} t_1} \sin \Omega_{\min} t_1 + D e^{-\frac{\gamma}{2} t_1} \cos \Omega_{\min} t_1 + \Bar{F}/K_{\min}}{A e^{\frac{\gamma}{2} t_0} \sin \Omega_{\max} t_0 - B e^{\frac{\gamma}{2} t_0} \cos \Omega_{\max} t_0 - \Bar{F}/K_{\max}}  
    \nonumber \\
    &\to \frac{(1/\Omega_{\min}) e^{-\frac{\gamma}{2} t_1^*} (\Omega_{\min}/\sqrt{K_{\min}/M}) }{(1/\Omega_{\max}) e^{\frac{\gamma}{2} t_0^*} (\Omega_{\max}/\sqrt{K_{\max}/M}) }  
    \nonumber \\
    &= \sqrt{\frac{K_{\max}}{K_{\min}}} e^{-\frac{\gamma}{2} (t_0^* + t_1^*)},
\end{align}
where we have used the equality: $\sin \left[\text{arccot}\left(-\frac{(\gamma/2)}{\Omega_{\max}}\right)\right] = \frac{\Omega_{\max}}{\sqrt{K_{\max}/M}}$, $\sin \left[\text{arccot}\frac{(\gamma/2)}{\Omega_{\min}}\right] = \frac{\Omega_{\min}}{\sqrt{K_{\min}/M}}$. Here $t_0^*$ and $t_1^*$ are the limiting values in Eq.~\ref{eq:appendix_t_0} and~\ref{eq:appendix_t_1}, respectively.

We define $\omega_{\max} = \frac{\Omega_{\max}}{(\gamma/2)} = \sqrt{\frac{K_{\max}}{M(\gamma/2)^2} - 1}$ and $\omega_{\min} = \frac{\Omega_{\min}}{(\gamma/2)} = \sqrt{\frac{K_{\min}}{M(\gamma/2)^2} - 1}$. After rearranging the terms, we simplifies the condition $\frac{q_1}{q_0}<1$ into
\begin{align}
    \frac{K_{\max}}{K_{\min}} < \exp\left[2\left(\frac{\text{arctan}\ \omega_{\min}}{\omega_{\min}} + \frac{\pi - \text{arctan}\ \omega_{\max}}{\omega_{\max}}\right)\right],
\end{align}
which is the inequality we see in Eq.~\ref{eq:error_convergence_condition} in Thm.~\ref{thm:appendix_error_convergence}. This completes the proof of Thm.~\ref{thm:appendix_error_convergence}. \qed

Note that the inequality in Eq.~\ref{eq:error_convergence_condition} in Thm.~\ref{thm:appendix_error_convergence} is satisfied in the limit $\frac{K_{\max}-K_{\min}}{K_{\min}}\to 0$. We can see this by take a log on both sides of the inequality and get
\begin{align}
    \log\left(1+\frac{K_{\max}-K_{\min}}{K_{\min}}\right) < 2\left(\frac{\text{arctan}\ \omega_{\min}}{\omega_{\min}} + \frac{\pi - \text{arctan}\ \omega_{\max}}{\omega_{\max}}\right),
\end{align}
whose left hand side is $O(\frac{K_{\max}-K_{\min}}{K_{\min}})$ and right hand side is $O(1)$ because $\omega_{\min}\approx\omega_{\max}$. Therefore, in the limit $\frac{K_{\max}-K_{\min}}{K_{\min}}\to 0$, the inequality always holds. This agrees with the Error Bounded Property of the approximately constant linear model stated in Prop.~\ref{prop:approx_const_linear_model}. (Actually the approximately constant linear model satisfies the Error Bounded Property under a more general condition: instead of $K>M(\gamma/2)^2$ we only need $K>0$ for the approximately constant linear model.)

If we transform the coordinates of the system to their normal modes, because the normal modes are also described by damped harmonic oscillators, all the analysis works the same for normal modes. This addresses the comment under Thm.~\ref{thm:error_convergence} in the main text.

\begin{figure}[t]
\centering
\includegraphics[width=8.5cm]{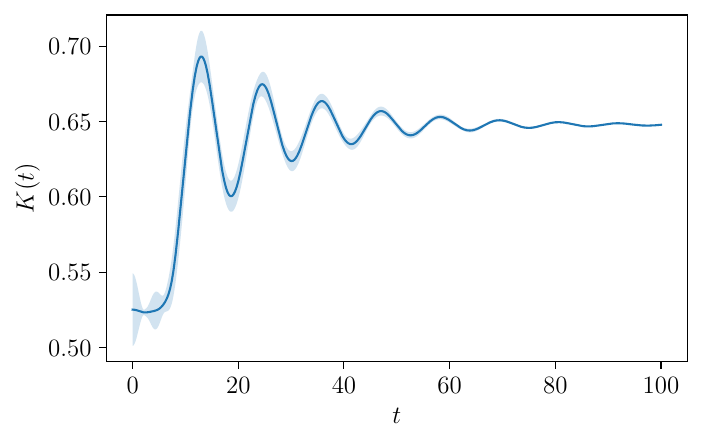}
\caption{Numerical measurement of the error stiffness $K(t)=k-g\frac{\partial n}{\partial Q}$ during the time range of interest. $g=1.4$ is used in the experiment. The highest point in the plot reaches 0.7104, and the lowest is 0.5011. Inserting these values into Eq.~\ref{eq:error_convergence_condition}, we get $LHS=1.418<1.501=RHS$.}
\label{fig:C_And}
\end{figure}

In Figure~\ref{fig:C_And}, we show an example of the fluctuation of the error stiffness $K(t)=k-g\frac{\partial n}{\partial Q}$ during the Holstein model dynamics. We see that the error stiffness $K(t)>M(\gamma/2)^2=0.0025$, where $M=1$ and $\gamma=0.1$. The highest and lowest points in the plot are 0.7104 and 0.5011. We insert these numbers into Eq.~\ref{eq:error_convergence_condition}. The left hand side of the condition results in 1.418, and the right hand side 1.501. Therefore, we confirm that the inequality in Eq.~\ref{eq:error_convergence_condition} in Thm.~\ref{thm:appendix_error_convergence} is satisfied during the time range we are interested in, and thus the Error Bounded Property is satisfied in the Holstein model.

\section{VI. Provably efficient adiabatic learning}
\label{sec:appendix_proof_PEAL}

Here we provide more details about how the PEAL algorithm is performed.

Step I: Collecting data. To be able to train a learning model, we first need to collect data samples from the distribution we are interested in. For example, one can sample a few different initial conditions, and use either classical solver or quantum computer to obtain a few dynamical evolution paths. The ``configuration''-``observable'' pairs sampled from those paths can serve as the data set we need. If one wants to predict other k-local observable that is not involved in the dynamics, one can compute that observable along the paths and collect the data for the next training step as well. 

Step II: ML training. A provable efficient ML algorithm for predicting ground state properties was presented recently~\cite{Preskill:2024nature}. Here we apply this algorithm to adiabatic dynamics. On the data set collected in Step I, a nonlinear feature map with geometrically local region information is performed. A model is trained with an $l_1$-regularized regression (LASSO) on the features. Hyperparameters in the model are properly chosen. One can train a single model for one type of observable if the system has translation symmetry. The same training process can be applied to the data set of observable that is not involved in the dynamics. Moreover, the learning model can be transferred to unseen physical parameter $g_\alpha$ in the Hamiltonian. An example is shown in our numerical experiment, also illustrated in Fig.~\ref{fig:single_path} in the main text.

Step III: ML prediction. For dynamical evolution paths with unseen initial conditions, we can use the PEAL prediction to speed up the simulation, instead of repeatedly using costly classical solver or quantum computer. The dynamical simulation is realized by iteratively updating the classical degree of freedom and the quantum degree of freedom. The classical updating step is done by classical ODE solver. The quantum updating step is done by ML prediction. In each iteration, we can add an extra correction step to achieve symmetry-preserving PEAL (see details in the following section). The provably efficiency and controllable error of the PEAL algorithm will be presented later in Thm.~\ref{thm:appendix_PEAL}.

Before we formally present Thm.~\ref{thm:appendix_PEAL}, we would like to mention two other theorems introduced in other works. In Ref.~\cite{Preskill:2024nature}, the authors prove two theorems:

\begin{theorem}
Consider any family of n-qubit geometrically-local Hamiltonians  
$\{H(x): x \in [-1,1]^m \}$ in a finite spatial dimension, such that each local term in $H(x)$ depends smoothly on $x$, and the smallest eigenvalue and the next smallest eigenvalues have a constant gap $\geq \Omega(1)$ between them. Then the ground state properties can be efficiently predicted. 
\end{theorem}

\begin{theorem}
\label{thm:appendix_N}
    
Given \( n, \delta > 0, \frac{1}{e} > \epsilon > 0 \) and a training data set \(\{x_t, y_t\}_{t=1}^N\) of size
\[
N = \log(n/\delta)2^{\text{polylog}(1/\epsilon)},
\]
where \(x_t\) is sampled from an unknown distribution \(D\) and \(|y_t - \text{Tr}(O \rho(x_t))| \leq \epsilon\) for any observable \(O\) with eigenvalues between \(-1\) and \(1\) that can be written as a sum of geometrically local observables. With a proper choice of the efficiently computable hyperparameters \(\delta_1, \delta_2\), and \(B\), the learned function \(h^*(x) = \mathbf{w}^* \cdot \phi(z)\) satisfies
\[
\mathbb{E}_{x \sim D} \left| h^*(z) - \text{Tr}(O \rho(z)) \right|^2 \leq \epsilon
\]
with probability at least \(1 - \delta\). The training and prediction time of the classical ML model are bounded by \(\mathcal{O}(nN) = n \log(n/\delta) 2^{\text{polylog}(1/\epsilon)}\).

\end{theorem}

The request in the first one, that the Hamiltonian has a constant gap, can be soften into that the correlation length has a finite upper bound.

Now we present the formal version of Thm.~\ref{thm:PEAL} in the main text.

\begin{theorem} [Provably Efficient Adiabatic Learning (PEAL) Theorem]
    \label{thm:appendix_PEAL}
    Given $n,\delta,\eta>0$, $\frac{1}{e}>\epsilon>0$ and a training data set $\{(g_\alpha, \Vec{P},\Vec{Q})_l,\langle\hat{O}_{\alpha,i}\rangle_l\}_{l=1}^N$ of size
    \begin{align}  
        N=\log(n/\delta)2^{\text{polylog}(1/\epsilon)},
        \label{eq:N}
    \end{align}
    where $l\in\{1,2,\dots,N\}$ is the index of data points, $(g_{\alpha})_l$ is the coupling constant used when collecting the data, $(\Vec{P},\Vec{Q})_l$ is classical variables sampled from exact simulation data, $\hat{O}_{\alpha,i}$ is an observable with eigenvalues in $[-1,1]$ that can be written as a sum of geometrically local observables, $\langle\hat{O}_{\alpha,i}\rangle_l$ is the ground state expectation value with an $n$-qubit gapped geometrically local Hamiltonian $\hat{H}((\Vec{P},\Vec{Q})_l)$. Apply an ML predicted model $\mathcal{M}$ learned with a proper choice of the efficiently computable hyperparameters. When the Error Bounded Property in Def.~\ref{def:error_bounded_property} is satisfied, a $T$-step PEAL prediction $Q_{i,\text{PEAL}}(T)$ comparing to the exact dynamical process $Q_{i,\text{Exact}}(T)$, i.e. the accumulated error, has an error bound
    \begin{align}
        |Q_{i,\text{PEAL}}(T) - Q_{i,\text{Exact}}(T)| \le O(\sqrt{T\epsilon/\eta})
        \label{eq:PEAL_bound_sqrt_T}
    \end{align}
    with probability $\mathbb{P} \ge 1-\eta-\delta$.

    If further assume the learning error of model $\mathcal{M}$ is sub-Gaussian distributed, the error bound can be improved into 
    \begin{align}
        |Q_{i,\text{PEAL}}(T) - Q_{i,\text{Exact}}(T)| \le O(\sqrt{2\epsilon\log(2T/\eta)})
        \label{eq:PEAL_bound_log_T}
    \end{align}
    with probability $\mathbb{P} \ge 1-\eta-\delta$.

    If further more assume the learning error of model $\mathcal{M}$ is bounded by $\sqrt{\epsilon}$ almost surely, the error bound can be further improved into
    \begin{align}
        |Q_{i,\text{PEAL}}(T) - Q_{i,\text{Exact}}(T)| \le O(\sqrt{\epsilon})
        \label{eq:PEAL_bound_const}
    \end{align}
    with probability $\mathbb{P} \ge 1-\delta$, i.e. the error is bounded by a constant.

    The accumulated error of classical variables $P_i$ and all $k$-local, bounded quantum observables also have the same scaling as the accumulated error of $Q_i$.
\end{theorem}

\textbf{Proof.} To be able to apply the Error Bounded Property in Def.~\ref{def:error_bounded_property}, we request that the prediction errors $|\delta\langle\hat{O}\rangle| = |\langle\hat{O}\rangle_{\text{PEAL}}(\Vec{P}_{\text{PEAL}},\Vec{Q}_{\text{PEAL}}) - \langle\hat{O}\rangle_{\text{Exact}}(\Vec{P}_{\text{PEAL}},\Vec{Q}_{\text{PEAL}})|$ are upper bounded for all $T$ number of time steps. From the Thm.~\ref{thm:appendix_N} above, we learn
\begin{align}
    \mathbb{E} |\delta\langle\hat{O}\rangle|^2 \le \epsilon,
\end{align}
from which we can estimate the probability of a single-shot prediction error going beyond a threshold:
\begin{align}
    \mathbb{P} \left[|\delta\langle\hat{O}\rangle| \ge A \sqrt{\epsilon}\right] &= \mathbb{P} \left[|\delta\langle\hat{O}\rangle|^2 \ge A^2 \epsilon\right] 
    \nonumber \\
    &\le \frac{\mathbb{E} |\delta\langle\hat{O}\rangle|^2}{A^2 \epsilon}
    \nonumber \\
    &\le \frac{1}{A^2},
\end{align}
where $A \ge 0$ is a constant factor we choose to describe the threshold. In the second line, we apply the Markov's inequality because $|\delta\langle\hat{O}\rangle|^2$ is non-negative.

For a dynamical simulation process with $T$ number of prediction stpes, the probability of any prediction error going beyond the threshold is controlled by the union bound:
\begin{align}
\label{eq:appendix_union_bound}
    \mathbb{P} \left[\bigcup_{i=1}^{T} \left\{|\delta\langle\hat{O}\rangle(t_i)| \ge A \sqrt{\epsilon} \right\}\right] \le \sum_{i=1}^{T} \mathbb{P} \left[|\delta\langle\hat{O}\rangle(t_i)| \ge A \sqrt{\epsilon}\right] \le \frac{T}{A^2}.
\end{align}

Taking into account that the trained model could have at most $\delta$ probability of failure, we get the probability of not being able to apply our PEAL method is at most ($\frac{T}{A^2} + \delta$) by another union bound. 

When we are able to control all the prediction errors under the threshold $A\sqrt{\epsilon}$, we have \begin{align}
    |q(t)|=|Q_{i,\text{PEAL}}(T) - Q_{i,\text{Exact}}(T)|\le O(A\sqrt{\epsilon})
\end{align}
by the Error Bounded Property in Def.~\ref{def:error_bounded_property}.  From the analysis above, we see this is applicable with a probability at least ($1 - \frac{T}{A^2} - \delta$).

For given $\eta>0$, we choose the threshold factor $A=\sqrt{T/\eta}$, such that $\eta = \frac{T}{A^2}$. Therefore, we get our most general bound
\begin{align}
    |Q_{i,\text{PEAL}}(T) - Q_{i,\text{Exact}}(T)|\le O(\sqrt{T\epsilon/\eta})
\label{eq:PEAL_bound_sqrt_T_proof}
\end{align}
with probability $\mathbb{P} \ge 1 - \eta - \delta$.

This ($1 - \eta - \delta$) behavior in the probability is expected, because we have an adiabatic hybrid algorithm with both quantum and classical updates. The $\delta$ failure comes from approximating the quantum algorithm, and the $\eta$ failure comes from the adiabatic quantum-classical dynamics.

This most general bound is showing a $\sqrt{T}$ behavior, which happens to be the same as a diffusion model. 

We can improve this bound if we can acquire more knowledge on the distribution of the prediction error $\delta\langle\hat{O}\rangle$. For example, if $\delta\langle\hat{O}\rangle$ has a sub-Gaussian distribution with a moment-generating function (MGF) 
\begin{align}
    M_{\delta\langle\hat{O}\rangle}(s) = \mathbb{E} \left[ e^{s\delta\langle\hat{O}\rangle} \right] \le \exp \frac{\epsilon s^2}{2},
\end{align}
then we can apply the Chernoff bound and get
\begin{align}
    \mathbb{P} \left[|\delta\langle\hat{O}\rangle| \ge B \sqrt{\epsilon}\right] \le 2\exp\left(-\frac{(B\sqrt{\epsilon})^2}{2\epsilon}\right) = 2 \exp \left(-B^2/2\right),
\end{align}
where $B$ is also a threshold factor.

Repeat the union bound argument, we can get $|Q_{i,\text{PEAL}}(T) - Q_{i,\text{Exact}}(T)|\le O(B\sqrt{\epsilon})$ with a probability at least ($1 - 2 T \exp \left(-B^2/2\right) - \delta$). For given $\eta>0$, we choose $B = \sqrt{2\log(2T/\eta)}$, such that $\eta = 2 T \exp \left(-B^2/2\right)$. Therefore, we get our bound with assuming sub-Gaussian prediction error
\begin{align}
    |Q_{i,\text{PEAL}}(T) - Q_{i,\text{Exact}}(T)|\le O(\sqrt{2\epsilon\log(2T/\eta)})
\label{eq:PEAL_bound_log_T_proof}
\end{align}
with probability $\mathbb{P} \ge 1 - \eta - \delta$.

This improves the general $\sqrt{T}$ behavior to a better $\sqrt{\log(T)}$ behavior.

We can improve this bound even further if we add even stronger assumption on the distribution of the prediction error $\delta\langle\hat{O}\rangle$. If we assume there exists a constant factor $C$ such that $|\delta\langle\hat{O}\rangle| \le C\sqrt{\epsilon}$ almost surely, then we can directly apply our Thm.~\ref{thm:appendix_error_convergence} and get
\begin{align}
    |Q_{i,\text{PEAL}}(T) - Q_{i,\text{Exact}}(T)|\le O(\sqrt{\epsilon})
\label{eq:PEAL_bound_const_proof}
\end{align}
with probability $\mathbb{P} \ge 1 - \delta$.

This is a constant bound which is independent on $T$. i.e. The accumulated position error of the dynamical simulation is bounded by a constant.

Because we also have a controlled $|p(t)|$ in Thm.~\ref{thm:appendix_error_convergence}, all the arguments above are also true if replacing $Q$ by $P$. 

With the proof of the following lemma, we complete the proof of Thm.~\ref{thm:appendix_PEAL}. \qed 

\begin{lemma}
\label{corollary:k_local_observables}
    All k-local, bounded observables have the same provably efficient bound under PEAL, no matter whether they are involved in the dynamics or not. For such an observable $\hat{\Omega}$, given a training data set $\{(\Vec{P},\Vec{Q})_l,\langle\hat{\Omega}\rangle_l\}_{l=1}^N$ of size $N$ same as in Thm.~\ref{thm:appendix_PEAL}, under various assumptions in Thm.~\ref{thm:appendix_PEAL}, the accumulated error $|\langle\hat{\Omega}\rangle_{\text{PEAL}}(\Vec{P}_{\text{PEAL}}(T),\Vec{Q}_{\text{PEAL}}(T)) - \langle\hat{\Omega}\rangle_{\text{Exact}}(\Vec{P}_{\text{Exact}}(T),\Vec{Q}_{\text{Exact}}(T))|$ holds similar bounds as in Eq.~\ref{eq:PEAL_bound_sqrt_T}, \ref{eq:PEAL_bound_log_T}, and~\ref{eq:PEAL_bound_const}.
\end{lemma}

\textbf{Proof.} The accumulated error for an observable is defined as 
\begin{align}
    |\langle\hat{\Omega}\rangle_{\text{PEAL}}(\Vec{P}_{\text{PEAL}}(T),\Vec{Q}_{\text{PEAL}}(T)) - \langle\hat{\Omega}\rangle_{\text{Exact}}(\Vec{P}_{\text{Exact}}(T),\Vec{Q}_{\text{Exact}}(T))|.
\end{align}

There are two source of error in this operator learning: one from the PEAL path $(\Vec{P}_{\text{PEAL}}(T),\Vec{Q}_{\text{PEAL}}(T))$ deviate from the exact simulation $(\Vec{P}_{\text{Exact}}(T),\Vec{Q}_{\text{Exact}}(T))$, the other from the single-step ML prediction error. In terms of formula, we have
\begin{align}
    &|\langle\hat{\Omega}\rangle_{\text{PEAL}}(\Vec{P}_{\text{PEAL}}(T),\Vec{Q}_{\text{PEAL}}(T)) - \langle\hat{\Omega}\rangle_{\text{Exact}}(\Vec{P}_{\text{Exact}}(T),\Vec{Q}_{\text{Exact}}(T))| \nonumber \\
    \le&\ |\langle\hat{\Omega}\rangle_{\text{PEAL}}(\Vec{P}_{\text{PEAL}}(T),\Vec{Q}_{\text{PEAL}}(T)) - \langle\hat{\Omega}\rangle_{\text{Exact}}(\Vec{P}_{\text{PEAL}}(T),\Vec{Q}_{\text{PEAL}}(T))|
    \nonumber \\
    &\ + |\langle\hat{\Omega}\rangle_{\text{Exact}}(\Vec{P}_{\text{PEAL}}(T),\Vec{Q}_{\text{PEAL}}(T)) - \langle\hat{\Omega}\rangle_{\text{Exact}}(\Vec{P}_{\text{Exact}}(T),\Vec{Q}_{\text{Exact}}(T))|.
\end{align}

We first analyze the second term by the perturbation theory. We have
\begin{align}
    \langle\hat{\Omega}\rangle_{\text{Exact}}(\Vec{P}_{\text{PEAL}}(T),\Vec{Q}_{\text{PEAL}}(T)) &= \langle \text{GS}_\text{PEAL}|\hat{\Omega}|\text{GS}_\text{PEAL}\rangle, 
    \\ 
    \langle\hat{\Omega}\rangle_{\text{Exact}}(\Vec{P}_{\text{Exact}}(T),\Vec{Q}_{\text{Exact}}(T)) &= \langle \text{GS}_\text{Exact}|\hat{\Omega}|\text{GS}_\text{Exact}\rangle, 
\end{align}
where $|\text{GS}_\text{PEAL/Exact}\rangle = |\text{GS}(\Vec{P}_{\text{PEAL/Exact}}(T),\Vec{Q}_{\text{PEAL/Exact}}(T))\rangle$ is the ground state wavefunction for the Hamiltonian with classical parameters $(\Vec{P}_{\text{PEAL/Exact}}(T),\Vec{Q}_{\text{PEAL/Exact}}(T))$, respectively.

Note that $(\Vec{P},\Vec{Q})$ are served as parameters of the Hamiltonian. For slightly different $(\Vec{P}_{\text{PEAL}}(T),\Vec{Q}_{\text{PEAL}}(T))$ and $(\Vec{P}_{\text{Exact}}(T),\Vec{Q}_{\text{Exact}}(T))$, the corresponding Hamiltonians are slightly different, and thus the difference between the ground states can be analyzed by first order perturbation
\begin{align}
\label{eq:appendix_GS_diff}
    |\text{GS}_\text{PEAL}\rangle - |\text{GS}_\text{Exact}\rangle = \sum_{n\ge1} \frac{\langle n_\text{Exact}|\hat{V}|\text{GS}_\text{Exact}\rangle}{E_\text{GS,Exact} - E_{n,\text{Exact}}} |n_\text{Exact}\rangle + o(\Vec{p}, \Vec{q}),
\end{align}
where $|n_\text{Exact}\rangle$ is the $n$-th excited state for the Hamiltonian with classical parameters $(\Vec{P}_{\text{Exact}}(T),\Vec{Q}_{\text{Exact}}(T))$. $E_\text{GS,Exact}$ and $E_{n,\text{Exact}}$ are energies of $|\text{GS}_\text{PEAL}\rangle$ and $|n_\text{Exact}\rangle$, respectively. We define the accumulated errors of classical parameters $\Vec{p}=\Vec{P}_{\text{PEAL}}(T)-\Vec{P}_{\text{Exact}}(T)$, $\Vec{q}=\Vec{Q}_{\text{PEAL}}(T)-\Vec{Q}_{\text{Exact}}(T)$. $\hat{V}$ is the first order Taylor expansion of the Hamiltonian
\begin{align}
    \hat{V} = \sum_i \left(\frac{\partial \hat{H}}{\partial Q_i} q_i + \frac{\partial \hat{H}}{\partial P_i} p_i\right).
\end{align}

Note the the dependence on $(\Vec{p}, \Vec{q})$ in Eq.~\ref{eq:appendix_GS_diff} only appears in $\hat{V}$, which is linear in $(\Vec{p}, \Vec{q})$. Moreover, Eq.~\ref{eq:appendix_GS_diff} is bounded by looking at its left hand side. Therefore, we see that overall $|\text{GS}_\text{PEAL}\rangle - |\text{GS}_\text{Exact}\rangle$ is of order $O(\Vec{p}, \Vec{q})$, and is controlled by the various bounds in Thm.~\ref{thm:appendix_PEAL}.

Therefore, the second term can be controlled by
\begin{align}
    & |\langle\hat{\Omega}\rangle_{\text{Exact}}(\Vec{P}_{\text{PEAL}}(T),\Vec{Q}_{\text{PEAL}}(T)) - \langle\hat{\Omega}\rangle_{\text{Exact}}(\Vec{P}_{\text{Exact}}(T),\Vec{Q}_{\text{Exact}}(T))| 
    \nonumber \\
    = &\ |\langle \text{GS}_\text{PEAL}|\hat{\Omega}|\text{GS}_\text{PEAL}\rangle - \langle \text{GS}_\text{Exact}|\hat{\Omega}|\text{GS}_\text{Exact}\rangle|
    \nonumber \\
    \le &\ |\langle \text{GS}_\text{PEAL}|\hat{\Omega}|\text{GS}_\text{PEAL}\rangle - \langle \text{GS}_\text{PEAL}|\hat{\Omega}|\text{GS}_\text{Exact}\rangle| + |\langle \text{GS}_\text{PEAL}|\hat{\Omega}|\text{GS}_\text{Exact}\rangle - \langle \text{GS}_\text{Exact}|\hat{\Omega}|\text{GS}_\text{Exact}\rangle|
    \nonumber \\
    = &\ |\langle \text{GS}_\text{PEAL}|\hat{\Omega} (|\text{GS}_\text{PEAL}\rangle - |\text{GS}_\text{Exact}\rangle )| + | (\langle \text{GS}_\text{PEAL}| - \langle \text{GS}_\text{Exact}| ) \hat{\Omega}|\text{GS}_\text{Exact}\rangle|,
\label{eq:appendix_second_term}
\end{align}
where $(|\text{GS}_\text{PEAL}\rangle - |\text{GS}_\text{Exact}\rangle )$ and $(\langle \text{GS}_\text{PEAL}| - \langle \text{GS}_\text{Exact}| )$ is of order $O(\Vec{p}, \Vec{q})$ and everything else is $O(1)$.
Therefore the last line in Eq.~\ref{eq:appendix_second_term} is of order $O(\Vec{p}, \Vec{q})$ because both terms are of order $O(\Vec{p}, \Vec{q})$. Hence, the second error term satisfies the same bounds as $Q$ does in Thm.~\ref{thm:appendix_PEAL}, which in the proof we analyze the probability of all $T$ steps' predictions having errors under a certain threshold.

The first term is the single-step prediction error $\delta\langle\hat{\Omega}\rangle$ at $T$-th step, which is analyzed in the proof of Thm.~\ref{thm:appendix_PEAL}. Combining the first term and the second term is equivalent to requiring ($T+1$) steps' prediction having errors under a certain threshold, which can be analyzed with the same technique we used in the proof of Thm.~\ref{thm:appendix_PEAL}. More precisely, we have ($T+1$) terms in the union bound in Eq.~\ref{eq:appendix_union_bound}. This is equivalent to having results in Thm.~\ref{thm:appendix_PEAL} by changing $T$ into ($T+1$). However, such change does not affect the scaling behaviors of the bounds with $T$.

Therefore, the first and second terms combined also satisfy the same bounds as $Q$ does in Thm.~\ref{thm:appendix_PEAL}. This completes the proof of Lemma~\ref{corollary:k_local_observables}, that the accumulated error of any $k$-local observable satisfy the same bounds in Thm.~\ref{thm:appendix_PEAL}. With this lemma proved, we complete the entire proof of Thm.~\ref{thm:appendix_PEAL}. \qed

\section{VII. PEAL Implementation}

Here we provide more details about how we implement the PEAL algorithm in our numerical experiment on the Holstein model.

PEAL Step I: Collecting data. For each of the six training values $g\in\{1.3, 1.32, 1.34, 1.36, 1.38, 1.4\}$, we randomly sampled 18 independent initial conditions $\Vec{Q}(t=0)\stackrel{\text{iid}}{\sim}\mathcal{N}(0,Q_v)^{\otimes L}$ with the standard deviation $Q_v=0.2$. $L=50$ is the system size. We set $\Vec{P}(t=0)=0$. For each initial condition, We iteratively perform quantum steps and classical steps, for 10,000 epochs, to generate a dynamical evolution path. We set the simulation time step $dt=0.01$ and the total time is 100. In the quantum steps, we use exact diagonalization (ED) to obtain the quantum observables. In the classical steps, we use fourth order Runge-Kutta method. For each path, we randomly sampled 500 pairs of $(\Vec{Q}_{\text{shift-}i}(t), n_i(t))$ data, where $t$ is a random time step and $i$ is a random site. $\Vec{Q}_{\text{shift-}i}$ is the $\Vec{Q}$ vector with elements rolled by a shift of $i$, so that the $i$-th element is put on the first place. Overall, we have $6\times18\times500=54,000$ data pairs, which effectively serve as 1,080 samples on the size $L=50$ system.

The purpose to use this kind of site-shifting data pairs is to respect the translation symmetry of the system. With periodic boundary condition, the system has translation symmetry. Therefore, we can use the same model to predict $n_i$ on any location in the system, as long as we also translate the $\Vec{Q}$ vector accordingly.

PEAL Step II: ML training. We use the provably efficient ML algorithm introduced in~\cite{Preskill:2024nature}. For each length-50 vector $\Vec{Q}$, we generate 50 local regions, each local region having radius 1 and consisting of 3 elements. We perform a random Fourier feature map on the local regions, generate $R$ number of cosine features and $R$ number of sine features for each local region. $R$ is a hyperparameter chosen from $\{5, 10, 20, 40, 80, 160\}$. The frequency factor $\gamma_\omega$ for the random Fourier feature map is another hyperparameter chosen from $\{0.3, 0.6, 1, 2, 3, 6, 10, 20\}$. We trained an $l_1$-regularized regression (LASSO) on the non-linear features. The strength of regularization $\alpha$ is determined by LassoLars method with 4-fold cross validation~\cite{sklearn}. For 1,024 samples, the grid-search best hyperparameters are $R=20$, $\gamma_\omega=6$, and $\alpha=4.98\times10^{-6}$. The best model has $316$ non-zero linear coefficients after the LASSO feature selection. 

PEAL Step III: ML prediction. To speed up the dynamical simulation, we replace the ED steps by the prediction from the learning model. We take care of the $U(1)$ global symmetry of the system. After each ML boosted quantum step, we subtract the electron density $n_i$ by the mean of their excess over $1/2$. Therefore, we can make the total electron number conserved at half filling, and respect the $U(1)$ global symmetry of the system. See more details in the following section.

\section{VIII. Details of Symmetry-preserving PEAL}

In the application of PEAL to specific systems with symmetries, we would like to preserve the symmetries during our numerical simulation. Here we present symmetry-preserving PEAL for two types of symmetries: The $U(1)$ global symmetry and the translation symmetry.

To preserve the $U(1)$ global symmetry, we would like to have the total charge conservation at each time step during the dynamics. When we sum up the model's prediction of the electron charge density at each site of the system, often case the result is not the same as the total charge in the previous time step. We can make a correction on the electron charge density at each site during each time step of the dynamics, in order to conserve the total charge.

There are two possible ways to do the correction. The first one is to uniformly add or subtract a constant on the electron charge density at each site. The second is to uniformly multiply a factor to the electron charge density at each site. However, the second way does not treat an electron (charge density equals to 1) and a hole (charge density equals to 0) equivalently. Therefore, we use the first way to do the correction.

Suppose the learning model $\mathcal{M}$ predicts the electron charge density $n_i=\langle c_i^\dagger c_i\rangle$ on each site $i$. In order to preserve the $U(1)$ global symmetry, we would like to have the corrected electron charge density $\Tilde{n}_i$ such that $\sum_i \Tilde{n}_i = L/2$, assuming at half filling for a system with $L$ sites. It is straightforward to see that we can achieve this charge conservation by doing the correction $n_i\to\Tilde{n}_i = n_i + \Delta n$ with $\Delta n = \frac{1}{L}\sum_i(\frac{1}{2}-n_i)$.

Now we prove the error bounds for our PEAL algorithm are still valid with this correction. Suppose each $n_i$ has a prediction error $\delta n_i=n_i-n_i^*$, where $n_i^*$ is the electron charge density computed with QSS, which satisfies $\sum_i n_i^* = L/2$. We can see $\Delta n=-\frac{1}{L}\sum_i \delta n_i$, and the error of the corrected electron charge density is $\delta\Tilde{n}_i = \Tilde{n}_i - n_i^* = \delta n_i + \Delta n = \delta n_i - \frac{1}{L}\sum_i \delta n_i$.

In the proof of PEAL error bounds, we analyzed all situations by turning the problem into the calculation of the probability of $|\delta\langle\hat{O}\rangle| \le D \sqrt{\epsilon}$, where the constant $D$ stands for $A$, $B$, or $C$ in the proof of Thm.~\ref{thm:appendix_PEAL}. Note that when all $|\delta n_i|\le D\sqrt{\epsilon}$, we have 
\begin{align}
    |\delta \Tilde{n}_i| = |\delta n_i - \frac{1}{L}\sum_i \delta n_i| \le |\delta n_i| + \frac{1}{L}\sum_i |\delta n_i| \le 2\, D \sqrt{\epsilon},
\end{align}
which means that we are safe to apply every error bounds after we raise the corresponding constant $D$ by a factor of $2$. This completes the proof.

When the system has a translation symmetry, such as the one for the periodic chain we studied in the main text, every site $i$ is on equal footing. Therefore, we will get the identical error distribution when we apply a model $\mathcal{M}_i$ trained for $n_i$ to another site $j$ to predict $n_j$. This means that we can save our cost by applying one model repeatedly on every site $i$ instead of training an independent model for every $n_i$, and we have the error bounds unaffected. Moreover, by doing so, we preserve the translation symmetry in the PEAL prediction: If instead of the original initial condition $\Vec{Q}(0)=(Q(0)_0, Q(0)_1, \dots, Q(0)_{L-2}, Q(0)_{L-1})$ we input a shifted initial condition $(Q(0)_1, Q(0)_2, \dots, Q(0)_{L-1}, Q(0)_{0})$, then all outputs from the symmetry-preserving PEAL algorithm will be also shifted correspondingly comparing to the original outputs.

\section{IX. Details of Numerical Experiments}

We are interested in the quantum dynamics of the Holstein model in Eq.~\ref{eq:H_ep} with a random initial distribution of $Q_i(0)$, and we set initial momentum $P_i(0)=0$. Under the adiabatic approximation, the quantum dynamics can be further described by three equations:  
\begin{align}
    Q^{'}_i(t) &= \frac{1}{M} P_i \\
    P^{'}_i(t) &= -k Q_i + g (n_i - \frac{1}{2}) - \gamma P_i \\
    H_e(\{Q_i(t)\}) &= -t \sum_{i,j} c_i^{\dagger} c_j - g \sum_i (c_i^{\dagger}c_i - \frac{1}{2}) Q_i
\end{align}
where $n_i = \langle c_i^{\dagger}c_i \rangle$ and $\gamma$ is the damping coefficient.

In the experiment, we observed three regimes of $g$ values. For a small $g$ value, the final stable configuration of the system has no CDW. This is due to the finite size effect. For the system size $L=50$ we used, the critical value to generate stable CDW is $g_{\text{crit}}=1.152$. This critical value can be analytically solved and we present the calculation already in earlier section. For a large $g$ value, domain walls can be generated in the system. The characteristic length of the domain walls decreases when $g$ increases. Therefore, when the system size is much larger than the domain wall characteristic length, domain walls can be relatively far from each other, which makes their interaction exponentially small and the equilibrium time exponentially large. For the system size $L=50$ we used, the value to generate long-lasting domain walls is $g_{\text{DW}}\sim 1.6$. For the value $g_{\text{crit}} < g < g_{\text{DW}}$, we can see a clear CDW developing during the dynamics. Therefore, we collect a data set for $g\in\{1.3, 1.32, 1.34, 1.36, 1.38, 1.4\}$, mimicking a uniform distribution in $[1.3,1.4]$.

For each $g$ value, we generate 18 ED dynamical evolution paths with random initial conditions, serving as the training data. We also generate 10 more ED dynamical evolution paths with random unseen initial conditions for each $g\in\{1.3, 1.32, 1.34, 1.36, 1.38, 1.4\}$ and for each $g\in\{1.31, 1.33, 1.35, 1.37, 1.39\}$, serving as the standard learning test set and the transfer learning test set, respectively. In Figure~\ref{fig:error_scaling} in the main text, we show how the test error scales with the number of samples used in model training. We trained models with number of samples from 1, 2, 4, ..., to 1,024. (Each sample contains 50 data pairs based on the system size we used.) We see an error scaling law agrees with the predicted logarithmic scaling. The transfer learning test error is only slightly higher than the standard learning test error. The scatter plots of the target values show the model has learned nicely for both the normal case and the transfer learning case.

\end{document}